% File jytex.tex, for jyTeX version 2.6M (June 1992)
% Copyright (c) 1991, 1992 by Jonathan P. Yamron
% For full documentation, "get jydoc" from hep-ph@xxx.lanl.gov
%   Problems?  Contact brahm@theory3.caltech.edu.

\catcode`\@=11

%*****************************************************************************

\message{Loading jyTeX fonts...}

%************************************************************
%*
%*             Available fonts
%*
%************************************************************

%************** 5-point fonts *******************************

\font\vptrm=cmr5 \font\vptmit=cmmi5 \font\vptsy=cmsy5 \font\vptbf=cmbx5

\skewchar\vptmit='177 \skewchar\vptsy='60 \fontdimen16
\vptsy=\the\fontdimen17 \vptsy

\def\vpt{\ifmmode\err@badsizechange\else
     \@mathfontinit
     \textfont0=\vptrm  \scriptfont0=\vptrm  \scriptscriptfont0=\vptrm
     \textfont1=\vptmit \scriptfont1=\vptmit \scriptscriptfont1=\vptmit
     \textfont2=\vptsy  \scriptfont2=\vptsy  \scriptscriptfont2=\vptsy
     \textfont3=\xptex  \scriptfont3=\xptex  \scriptscriptfont3=\xptex
     \textfont\bffam=\vptbf
     \scriptfont\bffam=\vptbf
     \scriptscriptfont\bffam=\vptbf
     \@fontstyleinit
     \def\rm{\vptrm\fam=\z@}%
     \def\bf{\vptbf\fam=\bffam}%
     \def\oldstyle{\vptmit\fam=\@ne}%
     \rm\fi}

%************** 6-point fonts *******************************

\font\viptrm=cmr6 \font\viptmit=cmmi6 \font\viptsy=cmsy6
\font\viptbf=cmbx6

\skewchar\viptmit='177 \skewchar\viptsy='60 \fontdimen16
\viptsy=\the\fontdimen17 \viptsy

\def\vipt{\ifmmode\err@badsizechange\else
     \@mathfontinit
     \textfont0=\viptrm  \scriptfont0=\vptrm  \scriptscriptfont0=\vptrm
     \textfont1=\viptmit \scriptfont1=\vptmit \scriptscriptfont1=\vptmit
     \textfont2=\viptsy  \scriptfont2=\vptsy  \scriptscriptfont2=\vptsy
     \textfont3=\xptex   \scriptfont3=\xptex  \scriptscriptfont3=\xptex
     \textfont\bffam=\viptbf
     \scriptfont\bffam=\vptbf
     \scriptscriptfont\bffam=\vptbf
     \@fontstyleinit
     \def\rm{\viptrm\fam=\z@}%
     \def\bf{\viptbf\fam=\bffam}%
     \def\oldstyle{\viptmit\fam=\@ne}%
     \rm\fi}
%************** 7-point fonts *******************************

\font\viiptrm=cmr7 \font\viiptmit=cmmi7 \font\viiptsy=cmsy7
\font\viiptit=cmti7 \font\viiptbf=cmbx7

\skewchar\viiptmit='177 \skewchar\viiptsy='60 \fontdimen16
\viiptsy=\the\fontdimen17 \viiptsy

\def\viipt{\ifmmode\err@badsizechange\else
     \@mathfontinit
     \textfont0=\viiptrm  \scriptfont0=\vptrm  \scriptscriptfont0=\vptrm
     \textfont1=\viiptmit \scriptfont1=\vptmit \scriptscriptfont1=\vptmit
     \textfont2=\viiptsy  \scriptfont2=\vptsy  \scriptscriptfont2=\vptsy
     \textfont3=\xptex    \scriptfont3=\xptex  \scriptscriptfont3=\xptex
     \textfont\itfam=\viiptit
     \scriptfont\itfam=\viiptit
     \scriptscriptfont\itfam=\viiptit
     \textfont\bffam=\viiptbf
     \scriptfont\bffam=\vptbf
     \scriptscriptfont\bffam=\vptbf
     \@fontstyleinit
     \def\rm{\viiptrm\fam=\z@}%
     \def\it{\viiptit\fam=\itfam}%
     \def\bf{\viiptbf\fam=\bffam}%
     \def\oldstyle{\viiptmit\fam=\@ne}%
     \rm\fi}

%************** 8-point fonts *******************************

\font\viiiptrm=cmr8 \font\viiiptmit=cmmi8 \font\viiiptsy=cmsy8
\font\viiiptit=cmti8
%\font\viiiptsl=cmsl8
\font\viiiptbf=cmbx8
%\font\viiipttt=cmtt8
%\font\viiiptss=cmss8

\skewchar\viiiptmit='177 \skewchar\viiiptsy='60 \fontdimen16
\viiiptsy=\the\fontdimen17 \viiiptsy

\def\viiipt{\ifmmode\err@badsizechange\else
     \@mathfontinit
     \textfont0=\viiiptrm  \scriptfont0=\viptrm  \scriptscriptfont0=\vptrm
     \textfont1=\viiiptmit \scriptfont1=\viptmit \scriptscriptfont1=\vptmit
     \textfont2=\viiiptsy  \scriptfont2=\viptsy  \scriptscriptfont2=\vptsy
     \textfont3=\xptex     \scriptfont3=\xptex   \scriptscriptfont3=\xptex
     \textfont\itfam=\viiiptit
     \scriptfont\itfam=\viiptit
     \scriptscriptfont\itfam=\viiptit
     \textfont\bffam=\viiiptbf
     \scriptfont\bffam=\viptbf
     \scriptscriptfont\bffam=\vptbf
     \@fontstyleinit
     \def\rm{\viiiptrm\fam=\z@}%
     \def\it{\viiiptit\fam=\itfam}%
     \def\bf{\viiiptbf\fam=\bffam}%
     \def\oldstyle{\viiiptmit\fam=\@ne}%
     \rm\fi}

%************** Optional 9-point fonts **********************

\def\getixpt{%
     \font\ixptrm=cmr9
     \font\ixptmit=cmmi9
     \font\ixptsy=cmsy9
     \font\ixptit=cmti9
%     \font\ixptsl=cmsl9
     \font\ixptbf=cmbx9
%     \font\ixpttt=cmtt9
%     \font\ixptss=cmss9
     \skewchar\ixptmit='177 \skewchar\ixptsy='60
     \fontdimen16 \ixptsy=\the\fontdimen17 \ixptsy}

\def\ixpt{\ifmmode\err@badsizechange\else
     \@mathfontinit
     \textfont0=\ixptrm  \scriptfont0=\viiptrm  \scriptscriptfont0=\vptrm
     \textfont1=\ixptmit \scriptfont1=\viiptmit \scriptscriptfont1=\vptmit
     \textfont2=\ixptsy  \scriptfont2=\viiptsy  \scriptscriptfont2=\vptsy
     \textfont3=\xptex   \scriptfont3=\xptex    \scriptscriptfont3=\xptex
     \textfont\itfam=\ixptit
     \scriptfont\itfam=\viiptit
     \scriptscriptfont\itfam=\viiptit
     \textfont\bffam=\ixptbf
     \scriptfont\bffam=\viiptbf
     \scriptscriptfont\bffam=\vptbf
     \@fontstyleinit
     \def\rm{\ixptrm\fam=\z@}%
     \def\it{\ixptit\fam=\itfam}%
     \def\bf{\ixptbf\fam=\bffam}%
     \def\oldstyle{\ixptmit\fam=\@ne}%
     \rm\fi}

%************** 10-point fonts ******************************

\font\xptrm=cmr10 \font\xptmit=cmmi10 \font\xptsy=cmsy10
\font\xptex=cmex10 \font\xptit=cmti10 \font\xptsl=cmsl10
\font\xptbf=cmbx10 \font\xpttt=cmtt10 \font\xptss=cmss10
\font\xptsc=cmcsc10 \font\xptbfs=cmb10 \font\xptbmit=cmmib10

\skewchar\xptmit='177 \skewchar\xptbmit='177 \skewchar\xptsy='60
\fontdimen16 \xptsy=\the\fontdimen17 \xptsy

\def\xpt{\ifmmode\err@badsizechange\else
     \@mathfontinit
     \textfont0=\xptrm  \scriptfont0=\viiptrm  \scriptscriptfont0=\vptrm
     \textfont1=\xptmit \scriptfont1=\viiptmit \scriptscriptfont1=\vptmit
     \textfont2=\xptsy  \scriptfont2=\viiptsy  \scriptscriptfont2=\vptsy
     \textfont3=\xptex  \scriptfont3=\xptex    \scriptscriptfont3=\xptex
     \textfont\itfam=\xptit
     \scriptfont\itfam=\viiptit
     \scriptscriptfont\itfam=\viiptit
     \textfont\bffam=\xptbf
     \scriptfont\bffam=\viiptbf
     \scriptscriptfont\bffam=\vptbf
     \textfont\bfsfam=\xptbfs
     \scriptfont\bfsfam=\viiptbf
     \scriptscriptfont\bfsfam=\vptbf
     \textfont\bmitfam=\xptbmit
     \scriptfont\bmitfam=\viiptmit
     \scriptscriptfont\bmitfam=\vptmit
     \@fontstyleinit
     \def\rm{\xptrm\fam=\z@}%
     \def\it{\xptit\fam=\itfam}%
     \def\sl{\xptsl}%
     \def\bf{\xptbf\fam=\bffam}%
     \def\tt{\xpttt}%
     \def\ss{\xptss}%
     \def\sc{\xptsc}%
     \def\bfs{\xptbfs\fam=\bfsfam}%
     \def\bmit{\fam=\bmitfam}%
     \def\oldstyle{\xptmit\fam=\@ne}%
     \rm\fi}

%************** Optional 11-point fonts *********************

\def\getxipt{%
     \font\xiptrm=cmr10  scaled\magstephalf
     \font\xiptmit=cmmi10 scaled\magstephalf
     \font\xiptsy=cmsy10 scaled\magstephalf
     \font\xiptex=cmex10 scaled\magstephalf
     \font\xiptit=cmti10 scaled\magstephalf
     \font\xiptsl=cmsl10 scaled\magstephalf
     \font\xiptbf=cmbx10 scaled\magstephalf
     \font\xipttt=cmtt10 scaled\magstephalf
     \font\xiptss=cmss10 scaled\magstephalf
     \skewchar\xiptmit='177 \skewchar\xiptsy='60
     \fontdimen16 \xiptsy=\the\fontdimen17 \xiptsy}

\def\xipt{\ifmmode\err@badsizechange\else
     \@mathfontinit
     \textfont0=\xiptrm  \scriptfont0=\viiiptrm  \scriptscriptfont0=\viptrm
     \textfont1=\xiptmit \scriptfont1=\viiiptmit \scriptscriptfont1=\viptmit
     \textfont2=\xiptsy  \scriptfont2=\viiiptsy  \scriptscriptfont2=\viptsy
     \textfont3=\xiptex  \scriptfont3=\xptex     \scriptscriptfont3=\xptex
     \textfont\itfam=\xiptit
     \scriptfont\itfam=\viiiptit
     \scriptscriptfont\itfam=\viiptit
     \textfont\bffam=\xiptbf
     \scriptfont\bffam=\viiiptbf
     \scriptscriptfont\bffam=\viptbf
     \@fontstyleinit
     \def\rm{\xiptrm\fam=\z@}%
     \def\it{\xiptit\fam=\itfam}%
     \def\sl{\xiptsl}%
     \def\bf{\xiptbf\fam=\bffam}%
     \def\tt{\xipttt}%
     \def\ss{\xiptss}%
     \def\oldstyle{\xiptmit\fam=\@ne}%
     \rm\fi}

%************** 12-point fonts ******************************

\font\xiiptrm=cmr12 \font\xiiptmit=cmmi12 \font\xiiptsy=cmsy10
scaled\magstep1 \font\xiiptex=cmex10  scaled\magstep1
\font\xiiptit=cmti12 \font\xiiptsl=cmsl12 \font\xiiptbf=cmbx12
%\font\xiipttt=cmtt12
\font\xiiptss=cmss12 \font\xiiptsc=cmcsc10 scaled\magstep1
\font\xiiptbfs=cmb10  scaled\magstep1 \font\xiiptbmit=cmmib10
scaled\magstep1

\skewchar\xiiptmit='177 \skewchar\xiiptbmit='177 \skewchar\xiiptsy='60
\fontdimen16 \xiiptsy=\the\fontdimen17 \xiiptsy

\def\xiipt{\ifmmode\err@badsizechange\else
     \@mathfontinit
     \textfont0=\xiiptrm  \scriptfont0=\viiiptrm  \scriptscriptfont0=\viptrm
     \textfont1=\xiiptmit \scriptfont1=\viiiptmit \scriptscriptfont1=\viptmit
     \textfont2=\xiiptsy  \scriptfont2=\viiiptsy  \scriptscriptfont2=\viptsy
     \textfont3=\xiiptex  \scriptfont3=\xptex     \scriptscriptfont3=\xptex
     \textfont\itfam=\xiiptit
     \scriptfont\itfam=\viiiptit
     \scriptscriptfont\itfam=\viiptit
     \textfont\bffam=\xiiptbf
     \scriptfont\bffam=\viiiptbf
     \scriptscriptfont\bffam=\viptbf
     \textfont\bfsfam=\xiiptbfs
     \scriptfont\bfsfam=\viiiptbf
     \scriptscriptfont\bfsfam=\viptbf
     \textfont\bmitfam=\xiiptbmit
     \scriptfont\bmitfam=\viiiptmit
     \scriptscriptfont\bmitfam=\viptmit
     \@fontstyleinit
     \def\rm{\xiiptrm\fam=\z@}%
     \def\it{\xiiptit\fam=\itfam}%
     \def\sl{\xiiptsl}%
     \def\bf{\xiiptbf\fam=\bffam}%
     \def\tt{\xiipttt}%
     \def\ss{\xiiptss}%
     \def\sc{\xiiptsc}%
     \def\bfs{\xiiptbfs\fam=\bfsfam}%
     \def\bmit{\fam=\bmitfam}%
     \def\oldstyle{\xiiptmit\fam=\@ne}%
     \rm\fi}

%************** Optional 13-point fonts *********************

\def\getxiiipt{%
     \font\xiiiptrm=cmr12  scaled\magstephalf
     \font\xiiiptmit=cmmi12 scaled\magstephalf
     \font\xiiiptsy=cmsy9  scaled\magstep2
     \font\xiiiptit=cmti12 scaled\magstephalf
     \font\xiiiptsl=cmsl12 scaled\magstephalf
     \font\xiiiptbf=cmbx12 scaled\magstephalf
     \font\xiiipttt=cmtt12 scaled\magstephalf
     \font\xiiiptss=cmss12 scaled\magstephalf
     \skewchar\xiiiptmit='177 \skewchar\xiiiptsy='60
     \fontdimen16 \xiiiptsy=\the\fontdimen17 \xiiiptsy}

\def\xiiipt{\ifmmode\err@badsizechange\else
     \@mathfontinit
     \textfont0=\xiiiptrm  \scriptfont0=\xptrm  \scriptscriptfont0=\viiptrm
     \textfont1=\xiiiptmit \scriptfont1=\xptmit \scriptscriptfont1=\viiptmit
     \textfont2=\xiiiptsy  \scriptfont2=\xptsy  \scriptscriptfont2=\viiptsy
     \textfont3=\xivptex   \scriptfont3=\xptex  \scriptscriptfont3=\xptex
     \textfont\itfam=\xiiiptit
     \scriptfont\itfam=\xptit
     \scriptscriptfont\itfam=\viiptit
     \textfont\bffam=\xiiiptbf
     \scriptfont\bffam=\xptbf
     \scriptscriptfont\bffam=\viiptbf
     \@fontstyleinit
     \def\rm{\xiiiptrm\fam=\z@}%
     \def\it{\xiiiptit\fam=\itfam}%
     \def\sl{\xiiiptsl}%
     \def\bf{\xiiiptbf\fam=\bffam}%
     \def\tt{\xiiipttt}%
     \def\ss{\xiiiptss}%
     \def\oldstyle{\xiiiptmit\fam=\@ne}%
     \rm\fi}

%************** 14-point fonts ******************************

\font\xivptrm=cmr12   scaled\magstep1 \font\xivptmit=cmmi12
scaled\magstep1 \font\xivptsy=cmsy10  scaled\magstep2
\font\xivptex=cmex10  scaled\magstep2 \font\xivptit=cmti12
scaled\magstep1 \font\xivptsl=cmsl12  scaled\magstep1
\font\xivptbf=cmbx12  scaled\magstep1
%\font\xivpttt=cmtt12  scaled\magstep1
\font\xivptss=cmss12  scaled\magstep1 \font\xivptsc=cmcsc10
scaled\magstep2 \font\xivptbfs=cmb10  scaled\magstep2
\font\xivptbmit=cmmib10 scaled\magstep2

\skewchar\xivptmit='177 \skewchar\xivptbmit='177 \skewchar\xivptsy='60
\fontdimen16 \xivptsy=\the\fontdimen17 \xivptsy

\def\xivpt{\ifmmode\err@badsizechange\else
     \@mathfontinit
     \textfont0=\xivptrm  \scriptfont0=\xptrm  \scriptscriptfont0=\viiptrm
     \textfont1=\xivptmit \scriptfont1=\xptmit \scriptscriptfont1=\viiptmit
     \textfont2=\xivptsy  \scriptfont2=\xptsy  \scriptscriptfont2=\viiptsy
     \textfont3=\xivptex  \scriptfont3=\xptex  \scriptscriptfont3=\xptex
     \textfont\itfam=\xivptit
     \scriptfont\itfam=\xptit
     \scriptscriptfont\itfam=\viiptit
     \textfont\bffam=\xivptbf
     \scriptfont\bffam=\xptbf
     \scriptscriptfont\bffam=\viiptbf
     \textfont\bfsfam=\xivptbfs
     \scriptfont\bfsfam=\xptbfs
     \scriptscriptfont\bfsfam=\viiptbf
     \textfont\bmitfam=\xivptbmit
     \scriptfont\bmitfam=\xptbmit
     \scriptscriptfont\bmitfam=\viiptmit
     \@fontstyleinit
     \def\rm{\xivptrm\fam=\z@}%
     \def\it{\xivptit\fam=\itfam}%
     \def\sl{\xivptsl}%
     \def\bf{\xivptbf\fam=\bffam}%
     \def\tt{\xivpttt}%
     \def\ss{\xivptss}%
     \def\sc{\xivptsc}%
     \def\bfs{\xivptbfs\fam=\bfsfam}%
     \def\bmit{\fam=\bmitfam}%
     \def\oldstyle{\xivptmit\fam=\@ne}%
     \rm\fi}

%************** 17-point fonts ******************************

\font\xviiptrm=cmr17 \font\xviiptmit=cmmi12 scaled\magstep2
\font\xviiptsy=cmsy10 scaled\magstep3 \font\xviiptex=cmex10
scaled\magstep3 \font\xviiptit=cmti12 scaled\magstep2
\font\xviiptbf=cmbx12 scaled\magstep2 \font\xviiptbfs=cmb10
scaled\magstep3

\skewchar\xviiptmit='177 \skewchar\xviiptsy='60 \fontdimen16
\xviiptsy=\the\fontdimen17 \xviiptsy

\def\xviipt{\ifmmode\err@badsizechange\else
     \@mathfontinit
     \textfont0=\xviiptrm  \scriptfont0=\xiiptrm  \scriptscriptfont0=\viiiptrm
     \textfont1=\xviiptmit \scriptfont1=\xiiptmit \scriptscriptfont1=\viiiptmit
     \textfont2=\xviiptsy  \scriptfont2=\xiiptsy  \scriptscriptfont2=\viiiptsy
     \textfont3=\xviiptex  \scriptfont3=\xiiptex  \scriptscriptfont3=\xptex
     \textfont\itfam=\xviiptit
     \scriptfont\itfam=\xiiptit
     \scriptscriptfont\itfam=\viiiptit
     \textfont\bffam=\xviiptbf
     \scriptfont\bffam=\xiiptbf
     \scriptscriptfont\bffam=\viiiptbf
     \textfont\bfsfam=\xviiptbfs
     \scriptfont\bfsfam=\xiiptbfs
     \scriptscriptfont\bfsfam=\viiiptbf
     \@fontstyleinit
     \def\rm{\xviiptrm\fam=\z@}%
     \def\it{\xviiptit\fam=\itfam}%
     \def\bf{\xviiptbf\fam=\bffam}%
     \def\bfs{\xviiptbfs\fam=\bfsfam}%
     \def\oldstyle{\xviiptmit\fam=\@ne}%
     \rm\fi}

%************** 21-point fonts ******************************

\font\xxiptrm=cmr17  scaled\magstep1
%\font\xxiptmit=cmmi12 scaled\magstep3
%\font\xxiptsy=cmsy10 scaled\magstep4
%\font\xxiptex=cmex10 scaled\magstep4
%\font\xxiptbf=cmbx12 scaled\magstep3

%\skewchar\xxiptmit='177 \skewchar\xxiptsy='60
%\fontdimen16 \xxiptsy=\the\fontdimen17 \xxiptsy

\def\xxipt{\ifmmode\err@badsizechange\else
     \@mathfontinit
%     \textfont0=\xxiptrm  \scriptfont0=\xivptrm  \scriptscriptfont0=\xptrm
%     \textfont1=\xxiptmit \scriptfont1=\xivptmit \scriptscriptfont1=\xptmit
%     \textfont2=\xxiptsy  \scriptfont2=\xivptsy  \scriptscriptfont2=\xptsy
%     \textfont3=\xxiptex  \scriptfont3=\xivptex  \scriptscriptfont3=\xptex
%     \textfont\bffam=\xxiptbf
%     \scriptfont\bffam=\xivptbf
%     \scriptscriptfont\bffam=\xptbf
     \@fontstyleinit
     \def\rm{\xxiptrm\fam=\z@}%
     \rm\fi}

%************** 25-point fonts ******************************

\font\xxvptrm=cmr17  scaled\magstep2
%\font\xxvptmit=cmmi12 scaled\magstep4
%\font\xxvptsy=cmsy10 scaled\magstep5
%\font\xxvptex=cmex10 scaled\magstep5
%\font\xxvptbf=cmbx12 scaled\magstep4

%\skewchar\xxvptmit='177 \skewchar\xxvptsy='60
%\fontdimen16 \xxvptsy=\the\fontdimen17 \xxvptsy

\def\xxvpt{\ifmmode\err@badsizechange\else
     \@mathfontinit
%     \textfont0=\xxvptrm  \scriptfont0=\xviiptrm  \scriptscriptfont0=\xiiptrm
%     \textfont1=\xxvptmit \scriptfont1=\xviiptmit \scriptscriptfont1=\xiiptmit
%     \textfont2=\xxvptsy  \scriptfont2=\xviiptsy  \scriptscriptfont2=\xiiptsy
%     \textfont3=\xxvptex  \scriptfont3=\xviiptex  \scriptscriptfont3=\xiiptex
%     \textfont\bffam=\xxvptbf
%     \scriptfont\bffam=\xviiptbf
%     \scriptscriptfont\bffam=\xiiptbf
     \@fontstyleinit
     \def\rm{\xxvptrm\fam=\z@}%
     \rm\fi}

%************** Other fonts *********************************

%\font\dummy=dummy

%******************************************************************************

\message{Loading jyTeX macros...}

%************************************************************
%*
%*              Simple modifications to plain
%*
%************************************************************
\message{modifications to plain.tex,}

% The "\outer" qualifier is removed from the definitions of \newcount through
% \newif so that they may be used in definitions.  \newif is also changed to
% make \if commands globally defined.

\def\newcount{\alloc@0\count\countdef\insc@unt}
\def\newdimen{\alloc@1\dimen\dimendef\insc@unt}
\def\newskip{\alloc@2\skip\skipdef\insc@unt}
\def\newmuskip{\alloc@3\muskip\muskipdef\@cclvi}
\def\newbox{\alloc@4\box\chardef\insc@unt}
\def\newtoks{\alloc@5\toks\toksdef\@cclvi}
\def\newhelp#1#2{\newtoks#1\global#1\expandafter{\csname#2\endcsname}}
\def\newread{\alloc@6\read\chardef\sixt@@n}
\def\newwrite{\alloc@7\write\chardef\sixt@@n}
\def\newfam{\alloc@8\fam\chardef\sixt@@n}
\def\newinsert#1{\global\advance\insc@unt by\m@ne
     \ch@ck0\insc@unt\count
     \ch@ck1\insc@unt\dimen
     \ch@ck2\insc@unt\skip
     \ch@ck4\insc@unt\box
     \allocationnumber=\insc@unt
     \global\chardef#1=\allocationnumber
     \wlog{\string#1=\string\insert\the\allocationnumber}}
\def\newif#1{\count@\escapechar \escapechar\m@ne
     \expandafter\expandafter\expandafter
          \xdef\@if#1{true}{\let\noexpand#1=\noexpand\iftrue}%
     \expandafter\expandafter\expandafter
          \xdef\@if#1{false}{\let\noexpand#1=\noexpand\iffalse}%
     \global\@if#1{false}\escapechar=\count@}

%************** Some parameter changes **********************

\newlinechar=`\^^J
\overfullrule=0pt

%************** Font-related modifications ******************

% The plain fonts are mapped onto the corresponding jyTeX fonts

% Some control sequences are disabled.

\let\itfam=\undefined

\let\bffam=\undefined

\count18=3

% German sharp s is given a new name (\ss is already taken)

\chardef\sharps="19

% The mathcode assignments of characters in the math italic font are changed to
% allow for switching to boldface.

\mathchardef\alpha="710B \mathchardef\beta="710C \mathchardef\gamma="710D
\mathchardef\delta="710E \mathchardef\epsilon="710F
\mathchardef\zeta="7110 \mathchardef\eta="7111 \mathchardef\theta="7112
\mathchardef\iota="7113 \mathchardef\kappa="7114
\mathchardef\lambda="7115 \mathchardef\mu="7116 \mathchardef\nu="7117
\mathchardef\xi="7118 \mathchardef\pi="7119 \mathchardef\rho="711A
\mathchardef\sigma="711B \mathchardef\tau="711C
\mathchardef\upsilon="711D \mathchardef\phi="711E \mathchardef\chi="711F
\mathchardef\psi="7120 \mathchardef\omega="7121
\mathchardef\varepsilon="7122 \mathchardef\vartheta="7123
\mathchardef\varpi="7124 \mathchardef\varrho="7125
\mathchardef\varsigma="7126 \mathchardef\varphi="7127
\mathchardef\imath="717B \mathchardef\jmath="717C \mathchardef\ell="7160
\mathchardef\wp="717D \mathchardef\partial="7140 \mathchardef\flat="715B
\mathchardef\natural="715C \mathchardef\sharp="715D

%************** Miscellaneous changes ***********************

% The dimension \p@ (1pt) is replaced with \rp@ (relative pt, defined below),
% whose size is determined by the base type size of the document.

\def\angle{{\vbox{\ialign{$\m@th\scriptstyle##$\crcr
     \not\mathrel{\mkern14mu}\crcr
     \noalign{\nointerlineskip}
     \mkern2.5mu\leaders\hrule height.34\rp@\hfill\mkern2.5mu\crcr}}}}
\def\vdots{\vbox{\baselineskip4\rp@ \lineskiplimit\z@
     \kern6\rp@\hbox{.}\hbox{.}\hbox{.}}}
\def\ddots{\mathinner{\mkern1mu\raise7\rp@\vbox{\kern7\rp@\hbox{.}}\mkern2mu
     \raise4\rp@\hbox{.}\mkern2mu\raise\rp@\hbox{.}\mkern1mu}}
\def\overrightarrow#1{\vbox{\ialign{##\crcr
     \rightarrowfill\crcr
     \noalign{\kern-\rp@\nointerlineskip}
     $\hfil\displaystyle{#1}\hfil$\crcr}}}
\def\overleftarrow#1{\vbox{\ialign{##\crcr
     \leftarrowfill\crcr
     \noalign{\kern-\rp@\nointerlineskip}
     $\hfil\displaystyle{#1}\hfil$\crcr}}}
\def\overbrace#1{\mathop{\vbox{\ialign{##\crcr
     \noalign{\kern3\rp@}
     \downbracefill\crcr
     \noalign{\kern3\rp@\nointerlineskip}
     $\hfil\displaystyle{#1}\hfil$\crcr}}}\limits}
\def\underbrace#1{\mathop{\vtop{\ialign{##\crcr
     $\hfil\displaystyle{#1}\hfil$\crcr
     \noalign{\kern3\rp@\nointerlineskip}
     \upbracefill\crcr
     \noalign{\kern3\rp@}}}}\limits}
\def\big#1{{\hbox{$\left#1\vbox to8.5\rp@ {}\right.\n@space$}}}
\def\Big#1{{\hbox{$\left#1\vbox to11.5\rp@ {}\right.\n@space$}}}
\def\bigg#1{{\hbox{$\left#1\vbox to14.5\rp@ {}\right.\n@space$}}}
\def\Bigg#1{{\hbox{$\left#1\vbox to17.5\rp@ {}\right.\n@space$}}}
\def\@vereq#1#2{\lower.5\rp@\vbox{\baselineskip\z@skip\lineskip-.5\rp@
     \ialign{$\m@th#1\hfil##\hfil$\crcr#2\crcr=\crcr}}}
\def\rlh@#1{\vcenter{\hbox{\ooalign{\raise2\rp@
     \hbox{$#1\rightharpoonup$}\crcr
     $#1\leftharpoondown$}}}}
\def\bordermatrix#1{\begingroup\m@th
     \setbox\z@\vbox{%
          \def\cr{\crcr\noalign{\kern2\rp@\global\let\cr\endline}}%
          \ialign{$##$\hfil\kern2\rp@\kern\p@renwd
               &\thinspace\hfil$##$\hfil&&\quad\hfil$##$\hfil\crcr
               \omit\strut\hfil\crcr
               \noalign{\kern-\baselineskip}%
               #1\crcr\omit\strut\cr}}%
     \setbox\tw@\vbox{\unvcopy\z@\global\setbox\@ne\lastbox}%
     \setbox\tw@\hbox{\unhbox\@ne\unskip\global\setbox\@ne\lastbox}%
     \setbox\tw@\hbox{$\kern\wd\@ne\kern-\p@renwd\left(\kern-\wd\@ne
          \global\setbox\@ne\vbox{\box\@ne\kern2\rp@}%
          \vcenter{\kern-\ht\@ne\unvbox\z@\kern-\baselineskip}%
          \,\right)$}%
     \null\;\vbox{\kern\ht\@ne\box\tw@}\endgroup}
\def\endinsert{\egroup
     \if@mid\dimen@\ht\z@
          \advance\dimen@\dp\z@
          \advance\dimen@12\rp@
          \advance\dimen@\pagetotal
          \ifdim\dimen@>\pagegoal\@midfalse\p@gefalse\fi
     \fi
     \if@mid\bigskip\box\z@
          \bigbreak
     \else\insert\topins{\penalty100 \splittopskip\z@skip
               \splitmaxdepth\maxdimen\floatingpenalty\z@
               \ifp@ge\dimen@\dp\z@
                    \vbox to\vsize{\unvbox\z@\kern-\dimen@}%
               \else\box\z@\nobreak\bigskip
               \fi}%
     \fi
     \endgroup}

% \normalbaselines is removed from \cases and \matrix.

\def\cases#1{\left\{\,\vcenter{\m@th
     \ialign{$##\hfil$&\quad##\hfil\crcr#1\crcr}}\right.}
\def\matrix#1{\null\,\vcenter{\m@th
     \ialign{\hfil$##$\hfil&&\quad\hfil$##$\hfil\crcr
          \mathstrut\crcr
          \noalign{\kern-\baselineskip}
          #1\crcr
          \mathstrut\crcr
          \noalign{\kern-\baselineskip}}}\,}

% \raggedbottom modified slightly

\newif\ifraggedbottom

\def\raggedbottom{\ifraggedbottom\else
     \advance\topskip by\z@ plus60pt \raggedbottomtrue\fi}%
\def\normalbottom{\ifraggedbottom
     \advance\topskip by\z@ plus-60pt \raggedbottomfalse\fi}

%************************************************************
%*
%*              Miscellaneous definitions
%*
%************************************************************
\message{hacks,}

%************** Hack registers ******************************

\toksdef\toks@i=1 \toksdef\toks@ii=2

%************** Basic macros ********************************

\def\TeX{T\kern-.1667em \lower.5ex \hbox{E}\kern-.125em X\null}
\def\jyTeX{{\leavevmode
     \raise.587ex \hbox{\it\j}\kern-.1em \lower.048ex \hbox{\it y}\kern-.12em
     \TeX}}

\let\then=\iftrue
\def\ifnoarg#1\then{\def\hack@{#1}\ifx\hack@\empty}
\def\ifundefined#1\then{%
     \expandafter\ifx\csname\expandafter\blank\string#1\endcsname\relax}
\def\useif#1\then{\csname#1\endcsname}
\def\usename#1{\csname#1\endcsname}
\def\useafter#1#2{\expandafter#1\csname#2\endcsname}

% Modify so that I can have \loop's within \loop's?
\long\def\loop#1\repeat{\def\@iterate{#1\expandafter\@iterate\fi}\@iterate
     \let\@iterate=\relax}
%\long\def\loop#1\repeat{\def\@loopbody{#1}\@iterate}
%\def\@iterate{\@loopbody\let\next=\@iterate\else\let\next=\relax\fi\next}

\let\TeXend=\end
\def\begin#1{\begingroup\def\@@blockname{#1}\usename{begin#1}}
\def\end#1{\usename{end#1}\def\hack@{#1}%
     \ifx\@@blockname\hack@
          \endgroup
     \else\err@badgroup\hack@\@@blockname
     \fi}
\def\@@blockname{}

\def\defaultoption[#1]#2{%
     \def\hack@{\ifx\hack@ii[\toks@={#2}\else\toks@={#2[#1]}\fi\the\toks@}%
     \futurelet\hack@ii\hack@}

\def\markup#1{\let\@@marksf=\empty
     \ifhmode\edef\@@marksf{\spacefactor=\the\spacefactor\relax}\/\fi
     ${}^{\hbox{\subscriptfonts#1}}$\@@marksf}

%************** Time registers ******************************

\newtoks\shortyear
\newtoks\militaryhour
\newtoks\standardhour
\newtoks\minute
\newtoks\amorpm

\def\settime{\count@=\time\divide\count@ by60
     \militaryhour=\expandafter{\number\count@}%
     {\multiply\count@ by-60 \advance\count@ by\time
          \xdef\hack@{\ifnum\count@<10 0\fi\number\count@}}%
     \minute=\expandafter{\hack@}%
     \ifnum\count@<12
          \amorpm={am}
     \else\amorpm={pm}
          \ifnum\count@>12 \advance\count@ by-12 \fi
     \fi
     \standardhour=\expandafter{\number\count@}%
     \def\hack@19##1##2{\shortyear={##1##2}}%
          \expandafter\hack@\the\year}

\def\monthword#1{%
     \ifcase#1
          $\bullet$\err@badcountervalue{monthword}%
          \or January\or February\or March\or April\or May\or June%
          \or July\or August\or September\or October\or November\or December%
     \else$\bullet$\err@badcountervalue{monthword}%
     \fi}

\def\monthabbr#1{%
     \ifcase#1
          $\bullet$\err@badcountervalue{monthabbr}%
          \or Jan\or Feb\or Mar\or Apr\or May\or Jun%
          \or Jul\or Aug\or Sep\or Oct\or Nov\or Dec%
     \else$\bullet$\err@badcountervalue{monthabbr}%
     \fi}

\def\militarytime{\the\militaryhour:\the\minute}
\def\standardtime{\the\standardhour:\the\minute}

%************** Number styles *******************************

\def\@setnumstyle#1#2{\expandafter\global\expandafter\expandafter
     \expandafter\let\expandafter\expandafter
     \csname @\expandafter\blank\string#1style\endcsname
     \csname#2\endcsname}
\def\numstyle#1{\usename{@\expandafter\blank\string#1style}#1}
\def\ifblank#1\then{\useafter\ifx{@\expandafter\blank\string#1}\blank}

\def\blank#1{}

\def\Roman#1{\expandafter\uppercase\expandafter{\romannumeral#1}}
\def\alphabetic#1{%
     \ifcase#1
          $\bullet$\err@badcountervalue{alphabetic}%
          \or a\or b\or c\or d\or e\or f\or g\or h\or i\or j\or k\or l\or m%
          \or n\or o\or p\or q\or r\or s\or t\or u\or v\or w\or x\or y\or z%
     \else$\bullet$\err@badcountervalue{alphabetic}%
     \fi}
\def\Alphabetic#1{\expandafter\uppercase\expandafter{\alphabetic{#1}}}
\def\symbols#1{%
     \ifcase#1
          $\bullet$\err@badcountervalue{symbols}%
          \or*\or\dag\or\ddag\or\S\or$\|$%
          \or**\or\dag\dag\or\ddag\ddag\or\S\S\or$\|\|$%
     \else$\bullet$\err@badcountervalue{symbols}%
     \fi}

%************** String macros *******************************

\catcode`\^^?=13 \def^^?{\relax}

\def\trimleading#1\to#2{\edef#2{#1}%
     \expandafter\@trimleading\expandafter#2#2^^?^^?}
\def\@trimleading#1#2#3^^?{\ifx#2^^?\def#1{}\else\def#1{#2#3}\fi}

\def\trimtrailing#1\to#2{\edef#2{#1}%
     \expandafter\@trimtrailing\expandafter#2#2^^? ^^?\relax}
\def\@trimtrailing#1#2 ^^?#3{\ifx#3\relax\toks@={}%
     \else\def#1{#2}\toks@={\trimtrailing#1\to#1}\fi
     \the\toks@}

\def\trim#1\to#2{\trimleading#1\to#2\trimtrailing#2\to#2}

\catcode`\^^?=15

%************** List macros *********************************

\long\def\additemL#1\to#2{\toks@={\^^\{#1}}\toks@ii=\expandafter{#2}%
     \xdef#2{\the\toks@\the\toks@ii}}

\long\def\additemR#1\to#2{\toks@={\^^\{#1}}\toks@ii=\expandafter{#2}%
     \xdef#2{\the\toks@ii\the\toks@}}

\def\getitemL#1\to#2{\expandafter\@getitemL#1\hack@#1#2}
\def\@getitemL\^^\#1#2\hack@#3#4{\def#4{#1}\def#3{#2}}

%************************************************************
%*
%*             Font-related macros
%*
%************************************************************
\message{font macros,}

%************** Font set-up *********************************

\newdimen\rp@
\newcount\@@sizeindex \@@sizeindex=0
\newcount\@@factori
\newcount\@@factorii
\newcount\@@factoriii
\newcount\@@factoriv

\countdef\maxfam=18
\newfam\itfam
\newfam\bffam
\newfam\bfsfam
\newfam\bmitfam

\def\@mathfontinit{\count@=4
     \loop\textfont\count@=\nullfont
          \scriptfont\count@=\nullfont
          \scriptscriptfont\count@=\nullfont
          \ifnum\count@<\maxfam\advance\count@ by\@ne
     \repeat}

\def\@fontstyleinit{%
     \def\it{\err@fontnotavailable\it}%
     \def\bf{\err@fontnotavailable\bf}%
     \def\bfs{\err@bfstobf}%
     \def\bmit{\err@fontnotavailable\bmit}%
     \def\sc{\err@fontnotavailable\sc}%
     \def\sl{\err@sltoit}%
     \def\ss{\err@fontnotavailable\ss}%
     \def\tt{\err@fontnotavailable\tt}}

\def\@parameterinit#1{\rm\rp@=.1em \@getscaling{#1}%
     \let\^^\=\@doscaling\scalingskipslist
     \setbox\strutbox=\hbox{\vrule
          height.708\baselineskip depth.292\baselineskip width\z@}}

\def\@getfactor#1#2#3#4{\@@factori=#1 \@@factorii=#2
     \@@factoriii=#3 \@@factoriv=#4}

\def\@getscaling#1{\count@=#1 \advance\count@ by-\@@sizeindex\@@sizeindex=#1
     \ifnum\count@<0
          \let\@mulordiv=\divide
          \let\@divormul=\multiply
          \multiply\count@ by\m@ne
     \else\let\@mulordiv=\multiply
          \let\@divormul=\divide
     \fi
     \edef\@@scratcha{\ifcase\count@                {1}{1}{1}{1}\or
          {1}{7}{23}{3}\or     {2}{5}{3}{1}\or      {9}{89}{13}{1}\or
          {6}{25}{6}{1}\or     {8}{71}{14}{1}\or    {6}{25}{36}{5}\or
          {1}{7}{53}{4}\or     {12}{125}{108}{5}\or {3}{14}{53}{5}\or
          {6}{41}{17}{1}\or    {13}{31}{13}{2}\or   {9}{107}{71}{2}\or
          {11}{139}{124}{3}\or {1}{6}{43}{2}\or     {10}{107}{42}{1}\or
          {1}{5}{43}{2}\or     {5}{69}{65}{1}\or    {11}{97}{91}{2}\fi}%
     \expandafter\@getfactor\@@scratcha}

\def\@doscaling#1{\@mulordiv#1by\@@factori\@divormul#1by\@@factorii
     \@mulordiv#1by\@@factoriii\@divormul#1by\@@factoriv}

%************* Size-changing commands ***********************

\newskip\headskip
\newskip\footskip

\def\typesize=#1pt{\count@=#1 \advance\count@ by-10
     \ifcase\count@
          \@setsizex\or\err@badtypesize\or
          \@setsizexii\or\err@badtypesize\or
          \@setsizexiv
     \else\err@badtypesize
     \fi}

\def\@setsizex{\getixpt
     \def\subsubscriptfonts{\vpt}%
          \def\subsubscriptsize{\vpt\@parameterinit{-8}}%
     \def\subscriptfonts{\viipt}\def\subscriptsize{\viipt\@parameterinit{-4}}%
     \def\footnotefonts{\viiipt}\def\footnotesize{\viiipt\@parameterinit{-2}}%
     \def\smallfonts{\ixpt}\def\smallsize{\ixpt\@parameterinit{-1}}%
     \def\normalfonts{\xpt}\def\normalsize{\xpt\@parameterinit{0}}%
     \def\bigfonts{\xiipt}\def\bigsize{\xiipt\@parameterinit{2}}%
     \def\Bigfonts{\xivpt}\def\Bigsize{\xivpt\@parameterinit{4}}%
     \def\biggfonts{\xviipt}\def\biggsize{\xviipt\@parameterinit{6}}%
     \def\Biggfonts{\xxipt}\def\Biggsize{\xxipt\@parameterinit{8}}%
     \def\tinyfonts{\vpt}\def\tinysize{\vpt\@parameterinit{-8}}%
     \def\HUGEFONTS{\xxvpt}\def\HUGESIZE{\xxvpt\@parameterinit{10}}%
     \normalsize\fixedskipslist}

\def\@setsizexii{\getxipt
     \def\subsubscriptfonts{\vipt}%
          \def\subsubscriptsize{\vipt\@parameterinit{-6}}%
     \def\subscriptfonts{\viiipt}%
          \def\subscriptsize{\viiipt\@parameterinit{-2}}%
     \def\footnotefonts{\xpt}\def\footnotesize{\xpt\@parameterinit{0}}%
     \def\smallfonts{\xipt}\def\smallsize{\xipt\@parameterinit{1}}%
     \def\normalfonts{\xiipt}\def\normalsize{\xiipt\@parameterinit{2}}%
     \def\bigfonts{\xivpt}\def\bigsize{\xivpt\@parameterinit{4}}%
     \def\Bigfonts{\xviipt}\def\Bigsize{\xviipt\@parameterinit{6}}%
     \def\biggfonts{\xxipt}\def\biggsize{\xxipt\@parameterinit{8}}%
     \def\Biggfonts{\xxvpt}\def\Biggsize{\xxvpt\@parameterinit{10}}%
     \def\tinyfonts{\vpt}\def\tinysize{\vpt\@parameterinit{-8}}%
     \def\HUGEFONTS{\xxvpt}\def\HUGESIZE{\xxvpt\@parameterinit{10}}%
     \normalsize\fixedskipslist}

\def\@setsizexiv{\getxiiipt
     \def\subsubscriptfonts{\viipt}%
          \def\subsubscriptsize{\viipt\@parameterinit{-4}}%
     \def\subscriptfonts{\xpt}\def\subscriptsize{\xpt\@parameterinit{0}}%
     \def\footnotefonts{\xiipt}\def\footnotesize{\xiipt\@parameterinit{2}}%
     \def\smallfonts{\xiiipt}\def\smallsize{\xiiipt\@parameterinit{3}}%
     \def\normalfonts{\xivpt}\def\normalsize{\xivpt\@parameterinit{4}}%
     \def\bigfonts{\xviipt}\def\bigsize{\xviipt\@parameterinit{6}}%
     \def\Bigfonts{\xxipt}\def\Bigsize{\xxipt\@parameterinit{8}}%
     \def\biggfonts{\xxvpt}\def\biggsize{\xxvpt\@parameterinit{10}}%
     \def\Biggfonts{\err@sizetoolarge\Biggfonts\HUGEFONTS}%
          \def\Biggsize{\err@sizetoolarge\Biggsize\HUGESIZE}%
     \def\tinyfonts{\vpt}\def\tinysize{\vpt\@parameterinit{-8}}%
     \def\HUGEFONTS{\xxvpt}\def\HUGESIZE{\xxvpt\@parameterinit{10}}%
     \normalsize\fixedskipslist}

\def\subsubscriptfonts{\vpt} \def\subsubscriptsize{\vpt\@parameterinit{-8}}
\def\subscriptfonts{\viipt}  \def\subscriptsize{\viipt\@parameterinit{-4}}
\def\footnotefonts{\viiipt}  \def\footnotesize{\viiipt\@parameterinit{-2}}
\def\smallfonts{\err@sizenotavailable\smallfonts}
                             \def\smallsize{\ixpt\@parameterinit{-1}}
\def\normalfonts{\xpt}       \def\normalsize{\xpt\@parameterinit{0}}
\def\bigfonts{\xiipt}        \def\bigsize{\xiipt\@parameterinit{2}}
\def\Bigfonts{\xivpt}        \def\Bigsize{\xivpt\@parameterinit{4}}
\def\biggfonts{\xviipt}      \def\biggsize{\xviipt\@parameterinit{6}}
\def\Biggfonts{\xxipt}       \def\Biggsize{\xxipt\@parameterinit{8}}
\def\tinyfonts{\vpt}         \def\tinysize{\vpt\@parameterinit{-8}}
\def\HUGEFONTS{\xxvpt}       \def\HUGESIZE{\xxvpt\@parameterinit{10}}

%************************************************************
%*
%*             Document layout
%*
%************************************************************
\message{document layout,}

%************** Page format *********************************

\newtoks\everyoutput \everyoutput={}
\newdimen\depthofpage
\newcount\pagenum \pagenum=0

\newdimen\oddtopmargin  \newdimen\eventopmargin
\newdimen\oddleftmargin \newdimen\evenleftmargin
\newtoks\oddhead        \newtoks\evenhead
\newtoks\oddfoot        \newtoks\evenfoot

\def\topmargin{\afterassignment\@seteventop\oddtopmargin}
\def\leftmargin{\afterassignment\@setevenleft\oddleftmargin}
\def\head{\afterassignment\@setevenhead\oddhead}
\def\foot{\afterassignment\@setevenfoot\oddfoot}

\def\@seteventop{\eventopmargin=\oddtopmargin}
\def\@setevenleft{\evenleftmargin=\oddleftmargin}
\def\@setevenhead{\evenhead=\oddhead}
\def\@setevenfoot{\evenfoot=\oddfoot}

\def\pagenumstyle#1{\@setnumstyle\pagenum{#1}}

\newif\ifdraft
\def\draft{\drafttrue\leftmargin=.5in \overfullrule=5pt }

\def\outputstyle#1{\global\expandafter\let\expandafter
          \@outputstyle\csname#1output\endcsname
     \usename{#1setup}}

\output={\@outputstyle}

\def\normaloutput{\the\everyoutput
     \global\advance\pagenum by\@ne
     \ifodd\pagenum
          \voffset=\oddtopmargin \hoffset=\oddleftmargin
     \else\voffset=\eventopmargin \hoffset=\evenleftmargin
     \fi
     \advance\voffset by-1in  \advance\hoffset by-1in
     \count0=\pagenum
     \expandafter\shipout\pagebox
     \ifnum\outputpenalty>-\@MM\else\dosupereject\fi}

\newdimen\fullhsize
\newbox\leftpage
\newcount\leftpagenum
\newcount\outputpagenum \outputpagenum=0
\let\leftorright=L

\def\twoupoutput{\the\everyoutput
     \global\advance\pagenum by\@ne
     \if L\leftorright
          \global\setbox\leftpage=\leftline{\pagebox}%
          \global\leftpagenum=\pagenum
          \global\let\leftorright=R%
     \else\global\advance\outputpagenum by\@ne
          \ifodd\outputpagenum
               \voffset=\oddtopmargin \hoffset=\oddleftmargin
          \else\voffset=\eventopmargin \hoffset=\evenleftmargin
          \fi
          \advance\voffset by-1in  \advance\hoffset by-1in
          \count0=\leftpagenum \count1=\pagenum
          \shipout\vbox{\hbox to\fullhsize
               {\box\leftpage\hfil\leftline{\pagebox}}}%
          \global\let\leftorright=L%
     \fi
     \ifnum\outputpenalty>-\@MM
     \else\dosupereject
          \if R\leftorright
               \globaldefs=\@ne\head={\hfil}\foot={\hfil}\globaldefs=\z@
               \null\newpage
          \fi
     \fi}

\def\pagebox{\vbox{\makeheadline\pagebody\makefootline}}

\def\makeheadline{%
     \vbox to\z@{\baselinestretch=\@m
          \vskip\topskip\vskip-.708\baselineskip\vskip-\headskip
          \line{\vbox to\ht\strutbox{}%
               \ifodd\pagenum\the\oddhead\else\the\evenhead\fi}%
          \vss}%
     \nointerlineskip}

\def\pagebody{\vbox to\vsize{%
     \boxmaxdepth\maxdepth
     \ifvoid\topins\else\unvbox\topins\fi
     \depthofpage=\dp255
     \unvbox255
     \ifraggedbottom\kern-\depthofpage\vfil\fi
     \ifvoid\footins
     \else\vskip\skip\footins
          \footnoterule
          \unvbox\footins
          \vskip-\footnoteskip
     \fi}}

\def\makefootline{\baselineskip=\footskip
     \line{\ifodd\pagenum\the\oddfoot\else\the\evenfoot\fi}}

%************** Sectioning commands *************************

\newskip\abovechapterskip
\newskip\belowchapterskip
\newskip\abovesectionskip
\newskip\belowsectionskip
\newskip\abovesubsectionskip
\newskip\belowsubsectionskip

\def\chapterstyle#1{\global\expandafter\let\expandafter\@chapterstyle
     \csname#1text\endcsname}
\def\sectionstyle#1{\global\expandafter\let\expandafter\@sectionstyle
     \csname#1text\endcsname}
\def\subsectionstyle#1{\global\expandafter\let\expandafter\@subsectionstyle
     \csname#1text\endcsname}

\def\chapter#1{%
     \ifdim\lastskip=17sp \else\chapterbreak\vskip\abovechapterskip\fi
     \@chapterstyle{\ifblank\chapternumstyle\then
          \else\newchapternum=\next\chapternumformat\ \fi#1}%
     \nobreak\vskip\belowchapterskip\vskip17sp }

\def\section#1{%
     \ifdim\lastskip=17sp \else\sectionbreak\vskip\abovesectionskip\fi
     \@sectionstyle{\ifblank\sectionnumstyle\then
          \else\newsectionnum=\next\sectionnumformat\ \fi#1}%
     \nobreak\vskip\belowsectionskip\vskip17sp }

\def\subsection#1{%
     \ifdim\lastskip=17sp \else\subsectionbreak\vskip\abovesubsectionskip\fi
     \@subsectionstyle{\ifblank\subsectionnumstyle\then
          \else\newsubsectionnum=\next\subsectionnumformat\ \fi#1}%
     \nobreak\vskip\belowsubsectionskip\vskip17sp }

%************** Text formatting commands ********************

\let\TeXunderline=\underline
\let\TeXoverline=\overline
\def\underline#1{\relax\ifmmode\TeXunderline{#1}\else
     $\TeXunderline{\hbox{#1}}$\fi}
\def\overline#1{\relax\ifmmode\TeXoverline{#1}\else
     $\TeXoverline{\hbox{#1}}$\fi}

\def\baselinestretch{\afterassignment\@baselinestretch\count@}
\def\@baselinestretch{\baselineskip=\normalbaselineskip
     \divide\baselineskip by\@m\baselineskip=\count@\baselineskip
     \setbox\strutbox=\hbox{\vrule
          height.708\baselineskip depth.292\baselineskip width\z@}%
     \bigskipamount=\the\baselineskip
          plus.25\baselineskip minus.25\baselineskip
     \medskipamount=.5\baselineskip
          plus.125\baselineskip minus.125\baselineskip
     \smallskipamount=.25\baselineskip
          plus.0625\baselineskip minus.0625\baselineskip}

\def\\{\ifhmode\ifnum\lastpenalty=-\@M\else\hfil\penalty-\@M\fi\fi
     \ignorespaces}
\def\newpage{\vfil\break}

\def\lefttext#1{\par{\@text\leftskip=\z@\rightskip=\centering
     \noindent#1\par}}
\def\righttext#1{\par{\@text\leftskip=\centering\rightskip=\z@
     \noindent#1\par}}
\def\centertext#1{\par{\@text\leftskip=\centering\rightskip=\centering
     \noindent#1\par}}
\def\@text{\parindent=\z@ \parfillskip=\z@ \everypar={}%
     \spaceskip=.3333em \xspaceskip=.5em
     \def\\{\ifhmode\ifnum\lastpenalty=-\@M\else\penalty-\@M\fi\fi
          \ignorespaces}}

\def\beginleft{\par\@text\leftskip=\z@ \rightskip=\centering}
     
\def\beginright{\par\@text\leftskip=\centering\rightskip=\z@ }
     
\def\begincenter{\par\@text\leftskip=\centering\rightskip=\centering}

\def\beginnarrow{\defaultoption[\parindent]\@beginnarrow}
\def\@beginnarrow[#1]{\par\advance\leftskip by#1\advance\rightskip by#1}

\begingroup
\catcode`\[=1 \catcode`\{=11 \gdef\beginignore[\endgroup\bgroup
     \catcode`\e=0 \catcode`\\=12 \catcode`\{=11 \catcode`\f=12 \let\or=\relax
     \let\nd{ignor=\fi \let\}=\egroup
     \iffalse}
\endgroup

\long\def\marginnote#1{\leavevmode
     \edef\@marginsf{\spacefactor=\the\spacefactor\relax}%
     \ifdraft\strut\vadjust{%
          \hbox to\z@{\hskip\hsize\hskip.1in
               \vbox to\z@{\vskip-\dp\strutbox
                    \marginnoteformat
                    \vskip-\ht\strutbox
                    \noindent\strut#1\par
                    \vss}%
               \hss}}%
     \fi
     \@marginsf}

%************** The \bye command ****************************

\newtoks\everybye \everybye={\par\vfil}
\outer\def\bye{\the\everybye
     \footnotecheck
     \prelabelcheck
     \streamcheck
     \supereject
     \TeXend}

%************************************************************
%*
%*             Footnotes
%*
%************************************************************
\message{footnotes,}

\newcount\footnotenum \footnotenum=0
\newskip\footnoteskip
\let\@footnotelist=\empty

\def\footnotenumstyle#1{\@setnumstyle\footnotenum{#1}%
     \useafter\ifx{@footnotenumstyle}\symbols
          \global\let\@footup=\empty
     \else\global\let\@footup=\markup
     \fi}

\def\footnote{\footnotecheck\defaultoption[]\@footnote}
\def\@footnote[#1]{\@footnotemark[#1]\@footnotetext}

\def\footnotemark{\defaultoption[]\@footnotemark}
\def\@footnotemark[#1]{\let\@footsf=\empty
     \ifhmode\edef\@footsf{\spacefactor=\the\spacefactor\relax}\/\fi
     \ifnoarg#1\then
          \global\advance\footnotenum by\@ne
          \@footup{\footnotenumformat}%
          \edef\@@foota{\footnotenum=\the\footnotenum\relax}%
          \expandafter\additemR\expandafter\@footup\expandafter
               {\@@foota\footnotenumformat}\to\@footnotelist
          \global\let\@footnotelist=\@footnotelist
     \else\markup{#1}%
          \additemR\markup{#1}\to\@footnotelist
          \global\let\@footnotelist=\@footnotelist
     \fi
     \@footsf}

\def\footnotetext{%
     \ifx\@footnotelist\empty\err@extrafootnotetext\else\@footnotetext\fi}
\def\@footnotetext{%
     \getitemL\@footnotelist\to\@@foota
     \global\let\@footnotelist=\@footnotelist
     \insert\footins\bgroup
     \footnoteformat
     \splittopskip=\ht\strutbox\splitmaxdepth=\dp\strutbox
     \interlinepenalty=\interfootnotelinepenalty\floatingpenalty=\@MM
     \noindent\llap{\@@foota}\strut
     \bgroup\aftergroup\@footnoteend
     \let\@@scratcha=}
\def\@footnoteend{\strut\par\vskip\footnoteskip\egroup}

\def\footnoterule{\normalfonts
     \kern-.3em \hrule width2in height.04em \kern .26em }

\def\footnotecheck{%
     \ifx\@footnotelist\empty
     \else\err@extrafootnotemark
          \global\let\@footnotelist=\empty
     \fi}

%************************************************************
%*
%*             Labelling macros
%*
%************************************************************
\message{labels,}

\let\@@labeldef=\xdef
\newif\if@labelfile
\newwrite\@labelfile
\let\@prelabellist=\empty

\def\label#1#2{\trim#1\to\@@labarg\edef\@@labtext{#2}%
     \edef\@@labname{lab@\@@labarg}%
     \useafter\ifundefined\@@labname\then\else\@yeslab\fi
     \useafter\@@labeldef\@@labname{#2}%
     \ifstreaming
          \expandafter\toks@\expandafter\expandafter\expandafter
               {\csname\@@labname\endcsname}%
          \immediate\write\streamout{\noexpand\label{\@@labarg}{\the\toks@}}%
     \fi}
\def\@yeslab{%
     \useafter\ifundefined{if\@@labname}\then
          \err@labelredef\@@labarg
     \else\useif{if\@@labname}\then
               \err@labelredef\@@labarg
          \else\global\usename{\@@labname true}%
               \useafter\ifundefined{pre\@@labname}\then
               \else\useafter\ifx{pre\@@labname}\@@labtext
                    \else\err@badlabelmatch\@@labarg
                    \fi
               \fi
               \if@labelfile
               \else\global\@labelfiletrue
                    \immediate\write\sixt@@n{--> Creating file \jobname.lab}%
                    \immediate\openout\@labelfile=\jobname.lab
               \fi
               \immediate\write\@labelfile
                    {\noexpand\prelabel{\@@labarg}{\@@labtext}}%
          \fi
     \fi}

\def\putlab#1{\trim#1\to\@@labarg\edef\@@labname{lab@\@@labarg}%
     \useafter\ifundefined\@@labname\then\@nolab\else\usename\@@labname\fi}
\def\@nolab{%
     \useafter\ifundefined{pre\@@labname}\then
          \undefinedlabelformat
          \err@needlabel\@@labarg
          \useafter\xdef\@@labname{\undefinedlabelformat}%
     \else\usename{pre\@@labname}%
          \useafter\xdef\@@labname{\usename{pre\@@labname}}%
     \fi
     \useafter\newif{if\@@labname}%
     \expandafter\additemR\@@labarg\to\@prelabellist}

\def\prelabel#1{\useafter\gdef{prelab@#1}}

\def\ifundefinedlabel#1\then{%
     \expandafter\ifx\csname lab@#1\endcsname\relax}
\def\useiflab#1\then{\csname iflab@#1\endcsname}

\def\prelabelcheck{{%
     \def\^^\##1{\useiflab{##1}\then\else\err@undefinedlabel{##1}\fi}%
     \@prelabellist}}

%************************************************************
%*
%*             Equation numbering
%*
%************************************************************
\message{equation numbering,}

\newcount\chapternum
\newcount\sectionnum
\newcount\subsectionnum
\newcount\equationnum
\newcount\subequationnum
\newcount\figurenum
\newcount\subfigurenum
\newcount\tablenum
\newcount\subtablenum

\newif\if@subeqncount
\newif\if@subfigcount
\newif\if@subtblcount

\def\newchapternum{\newsectionnum=\z@\@resetnum\chapternum}
\def\newsectionnum{\newsubsectionnum=\z@\@resetnum\sectionnum}
\def\newsubsectionnum{\newequationnum=\z@\newfigurenum=\z@\newtablenum=\z@
     \@resetnum\subsectionnum}
\def\newequationnum{\newsubequationnum=\z@\@resetnum\equationnum}
\def\newsubequationnum{\@resetnum\subequationnum}
\def\newfigurenum{\newsubfigurenum=\z@\@resetnum\figurenum}
\def\newsubfigurenum{\@resetnum\subfigurenum}
\def\newtablenum{\newsubtablenum=\z@\@resetnum\tablenum}
\def\newsubtablenum{\@resetnum\subtablenum}

\def\@resetnum#1{\global\advance#1by1 \edef\next{\the#1\relax}\global#1}

\newchapternum=0

\def\chapternumstyle#1{\@setnumstyle\chapternum{#1}}
\def\sectionnumstyle#1{\@setnumstyle\sectionnum{#1}}
\def\subsectionnumstyle#1{\@setnumstyle\subsectionnum{#1}}
\def\equationnumstyle#1{\@setnumstyle\equationnum{#1}}
\def\subequationnumstyle#1{\@setnumstyle\subequationnum{#1}%
     \ifblank\subequationnumstyle\then\global\@subeqncountfalse\fi
     \ignorespaces}
\def\figurenumstyle#1{\@setnumstyle\figurenum{#1}}
\def\subfigurenumstyle#1{\@setnumstyle\subfigurenum{#1}%
     \ifblank\subfigurenumstyle\then\global\@subfigcountfalse\fi
     \ignorespaces}
\def\tablenumstyle#1{\@setnumstyle\tablenum{#1}}
\def\subtablenumstyle#1{\@setnumstyle\subtablenum{#1}%
     \ifblank\subtablenumstyle\then\global\@subtblcountfalse\fi
     \ignorespaces}

\def\eqnlabel#1{%
     \if@subeqncount
          \newsubequationnum=\next
     \else\newequationnum=\next
          \ifblank\subequationnumstyle\then
          \else\global\@subeqncounttrue
               \newsubequationnum=\@ne
          \fi
     \fi
     \label{#1}{\puteqnformat}(\puteqn{#1})%
     \ifdraft\rlap{\hskip.1in{\tt#1}}\fi}

\let\puteqn=\putlab

\def\equation#1#2{\useafter\gdef{eqn@#1}{#2\eqno\eqnlabel{#1}}}
\def\Equation#1{\useafter\gdef{eqn@#1}}

\def\putequation#1{\useafter\ifundefined{eqn@#1}\then
     \err@undefinedeqn{#1}\else\usename{eqn@#1}\fi}

\def\eqnseriesstyle#1{\gdef\@eqnseriesstyle{#1}}
\def\begineqnseries{\subequationnumstyle{\@eqnseriesstyle}%
     \defaultoption[]\@begineqnseries}
\def\@begineqnseries[#1]{\edef\@@eqnname{#1}}
\def\endeqnseries{\subequationnumstyle{blank}%
     \expandafter\ifnoarg\@@eqnname\then
     \else\label\@@eqnname{\puteqnformat}%
     \fi
     \aftergroup\ignorespaces}

\def\figlabel#1{%
     \if@subfigcount
          \newsubfigurenum=\next
     \else\newfigurenum=\next
          \ifblank\subfigurenumstyle\then
          \else\global\@subfigcounttrue
               \newsubfigurenum=\@ne
          \fi
     \fi
     \label{#1}{\putfigformat}\putfig{#1}%
     {\def\marginnoteformat{\tt}\marginnote{#1}}}

\let\putfig=\putlab

\def\figseriesstyle#1{\gdef\@figseriesstyle{#1}}
\def\beginfigseries{\subfigurenumstyle{\@figseriesstyle}%
     \defaultoption[]\@beginfigseries}
\def\@beginfigseries[#1]{\edef\@@figname{#1}}
\def\endfigseries{\subfigurenumstyle{blank}%
     \expandafter\ifnoarg\@@figname\then
     \else\label\@@figname{\putfigformat}%
     \fi
     \aftergroup\ignorespaces}

\def\tbllabel#1{%
     \if@subtblcount
          \newsubtablenum=\next
     \else\newtablenum=\next
          \ifblank\subtablenumstyle\then
          \else\global\@subtblcounttrue
               \newsubtablenum=\@ne
          \fi
     \fi
     \label{#1}{\puttblformat}\puttbl{#1}%
     {\def\marginnoteformat{\tt}\marginnote{#1}}}

\let\puttbl=\putlab

\def\tblseriesstyle#1{\gdef\@tblseriesstyle{#1}}
\def\begintblseries{\subtablenumstyle{\@tblseriesstyle}%
     \defaultoption[]\@begintblseries}
\def\@begintblseries[#1]{\edef\@@tblname{#1}}
\def\endtblseries{\subtablenumstyle{blank}%
     \expandafter\ifnoarg\@@tblname\then
     \else\label\@@tblname{\puttblformat}%
     \fi
     \aftergroup\ignorespaces}

%************************************************************
%*
%*             Reference numbering
%*
%************************************************************
\message{reference numbering,}

\newcount\referencenum \referencenum=0
\newcount\@@prerefcount \@@prerefcount=0
\newcount\@@thisref
\newcount\@@lastref
\newcount\@@loopref
\newcount\@@refseq
\newdimen\refnumindent
\let\@undefreflist=\empty

\def\referencenumstyle#1{\@setnumstyle\referencenum{#1}}

\def\referencestyle#1{\usename{@ref#1}}

\def\@refsequential{%
     \gdef\@refpredef##1{\global\advance\referencenum by\@ne
          \let\^^\=0\label{##1}{\^^\{\the\referencenum}}%
          \useafter\gdef{ref@\the\referencenum}{{##1}{\undefinedlabelformat}}}%
     \gdef\@reference##1##2{%
          \ifundefinedlabel##1\then
          \else\def\^^\####1{\global\@@thisref=####1\relax}\putlab{##1}%
               \useafter\gdef{ref@\the\@@thisref}{{##1}{##2}}%
          \fi}%
     \gdef\endputreferences{%
          \loop\ifnum\@@loopref<\referencenum
                    \advance\@@loopref by\@ne
                    \expandafter\expandafter\expandafter\@printreference
                         \csname ref@\the\@@loopref\endcsname
          \repeat
          \par}}

\def\@refpreordered{%
     \gdef\@refpredef##1{\global\advance\referencenum by\@ne
          \additemR##1\to\@undefreflist}%
     \gdef\@reference##1##2{%
          \ifundefinedlabel##1\then
          \else\global\advance\@@loopref by\@ne
               {\let\^^\=0\label{##1}{\^^\{\the\@@loopref}}}%
               \@printreference{##1}{##2}%
          \fi}
     \gdef\endputreferences{%
          \def\^^\####1{\useiflab{####1}\then
               \else\reference{####1}{\undefinedlabelformat}\fi}%
          \@undefreflist
          \par}}

\def\beginprereferences{\par
     \def\reference##1##2{\global\advance\referencenum by1\@ne
          \let\^^\=0\label{##1}{\^^\{\the\referencenum}}%
          \useafter\gdef{ref@\the\referencenum}{{##1}{##2}}}}
\def\endprereferences{\global\@@prerefcount=\the\referencenum\par}

\def\beginputreferences{\par
     \refnumindent=\z@\@@loopref=\z@
     \loop\ifnum\@@loopref<\referencenum
               \advance\@@loopref by\@ne
               \setbox\z@=\hbox{\referencenum=\@@loopref
                    \referencenumformat\enskip}%
               \ifdim\wd\z@>\refnumindent\refnumindent=\wd\z@\fi
     \repeat
     \putreferenceformat
     \@@loopref=\z@
     \loop\ifnum\@@loopref<\@@prerefcount
               \advance\@@loopref by\@ne
               \expandafter\expandafter\expandafter\@printreference
                    \csname ref@\the\@@loopref\endcsname
     \repeat
     \let\reference=\@reference}

\def\@printreference#1#2{\ifx#2\undefinedlabelformat\err@undefinedref{#1}\fi
     \noindent\ifdraft\rlap{\hskip\hsize\hskip.1in \tt#1}\fi
     \llap{\referencenum=\@@loopref\referencenumformat\enskip}#2\par}

\def\reference#1#2{{\par\refnumindent=\z@\putreferenceformat\noindent#2\par}}

\def\putref#1{\trim#1\to\@@refarg
     \expandafter\ifnoarg\@@refarg\then
          \toks@={\relax}%
     \else\@@lastref=-\@m\def\@@refsep{}\def\@more{\@nextref}%
          \toks@={\@nextref#1,,}%
     \fi\the\toks@}
\def\@nextref#1,{\trim#1\to\@@refarg
     \expandafter\ifnoarg\@@refarg\then
          \let\@more=\relax
     \else\ifundefinedlabel\@@refarg\then
               \expandafter\@refpredef\expandafter{\@@refarg}%
          \fi
          \def\^^\##1{\global\@@thisref=##1\relax}%
          \global\@@thisref=\m@ne
          \setbox\z@=\hbox{\putlab\@@refarg}%
     \fi
     \advance\@@lastref by\@ne
     \ifnum\@@lastref=\@@thisref\advance\@@refseq by\@ne\else\@@refseq=\@ne\fi
     \ifnum\@@lastref<\z@
     \else\ifnum\@@refseq<\thr@@
               \@@refsep\def\@@refsep{,}%
               \ifnum\@@lastref>\z@
                    \advance\@@lastref by\m@ne
                    {\referencenum=\@@lastref\putrefformat}%
               \else\undefinedlabelformat
               \fi
          \else\def\@@refsep{--}%
          \fi
     \fi
     \@@lastref=\@@thisref
     \@more}

%************************************************************
%*
%*             Job streaming
%*
%************************************************************
\message{streaming,}

\newif\ifstreaming

\def\streamto{\defaultoption[\jobname]\@streamto}
\def\@streamto[#1]{\global\streamingtrue
     \immediate\write\sixt@@n{--> Streaming to #1.str}%
     \newwrite\streamout\immediate\openout\streamout=#1.str }

\def\streamfrom{\defaultoption[\jobname]\@streamfrom}
\def\@streamfrom[#1]{\newread\streamin\openin\streamin=#1.str
     \ifeof\streamin
          \expandafter\err@nostream\expandafter{#1.str}%
     \else\immediate\write\sixt@@n{--> Streaming from #1.str}%
          \let\@@labeldef=\gdef
          \ifstreaming
               \edef\@elc{\endlinechar=\the\endlinechar}%
               \endlinechar=\m@ne
               \loop\read\streamin to\@@scratcha
                    \ifeof\streamin
                         \streamingfalse
                    \else\toks@=\expandafter{\@@scratcha}%
                         \immediate\write\streamout{\the\toks@}%
                    \fi
                    \ifstreaming
               \repeat
               \@elc
               \input #1.str
               \streamingtrue
          \else\input #1.str
          \fi
          \let\@@labeldef=\xdef
     \fi}

\def\streamcheck{\ifstreaming
     \immediate\write\streamout{\pagenum=\the\pagenum}%
     \immediate\write\streamout{\footnotenum=\the\footnotenum}%
     \immediate\write\streamout{\referencenum=\the\referencenum}%
     \immediate\write\streamout{\chapternum=\the\chapternum}%
     \immediate\write\streamout{\sectionnum=\the\sectionnum}%
     \immediate\write\streamout{\subsectionnum=\the\subsectionnum}%
     \immediate\write\streamout{\equationnum=\the\equationnum}%
     \immediate\write\streamout{\subequationnum=\the\subequationnum}%
     \immediate\write\streamout{\figurenum=\the\figurenum}%
     \immediate\write\streamout{\subfigurenum=\the\subfigurenum}%
     \immediate\write\streamout{\tablenum=\the\tablenum}%
     \immediate\write\streamout{\subtablenum=\the\subtablenum}%
     \immediate\closeout\streamout
     \fi}

%************************************************************
%*
%*             Error messages
%*
%************************************************************

\def\err@badtypesize{%
     \errhelp={The limited availability of certain fonts requires^^J%
          that the base type size be 10pt, 12pt, or 14pt.^^J}%
     \errmessage{--> Illegal base type size}}

\def\err@badsizechange{\immediate\write\sixt@@n
     {--> Size change not allowed in math mode, ignored}}

\def\err@sizetoolarge#1{\immediate\write\sixt@@n
     {--> \noexpand#1 too big, substituting HUGE}}

\def\err@sizenotavailable#1{\immediate\write\sixt@@n
     {--> Size not available, \noexpand#1 ignored}}

\def\err@fontnotavailable#1{\immediate\write\sixt@@n
     {--> Font not available, \noexpand#1 ignored}}

\def\err@sltoit{\immediate\write\sixt@@n
     {--> Style \noexpand\sl not available, substituting \noexpand\it}%
     \it}

\def\err@bfstobf{\immediate\write\sixt@@n
     {--> Style \noexpand\bfs not available, substituting \noexpand\bf}%
     \bf}

\def\err@badgroup#1#2{%
     \errhelp={The block you have just tried to close was not the one^^J%
          most recently opened.^^J}%
     \errmessage{--> \noexpand\end{#1} doesn't match \noexpand\begin{#2}}}

\def\err@badcountervalue#1{\immediate\write\sixt@@n
     {--> Counter (#1) out of bounds}}

\def\err@extrafootnotemark{\immediate\write\sixt@@n
     {--> \noexpand\footnotemark command
          has no corresponding \noexpand\footnotetext}}

\def\err@extrafootnotetext{%
     \errhelp{You have given a \noexpand\footnotetext command without first
          specifying^^Ja \noexpand\footnotemark.^^J}%
     \errmessage{--> \noexpand\footnotetext command has no corresponding
          \noexpand\footnotemark}}

\def\err@labelredef#1{\immediate\write\sixt@@n
     {--> Label "#1" redefined}}

\def\err@badlabelmatch#1{\immediate\write\sixt@@n
     {--> Definition of label "#1" doesn't match value in \jobname.lab}}

\def\err@needlabel#1{\immediate\write\sixt@@n
     {--> Label "#1" cited before its definition}}

\def\err@undefinedlabel#1{\immediate\write\sixt@@n
     {--> Label "#1" cited but never defined}}

\def\err@undefinedeqn#1{\immediate\write\sixt@@n
     {--> Equation "#1" not defined}}

\def\err@undefinedref#1{\immediate\write\sixt@@n
     {--> Reference "#1" not defined}}

\def\err@nostream#1{%
     \errhelp={You have tried to input a stream file that doesn't exist.^^J}%
     \errmessage{--> Stream file #1 not found}}

%************************************************************
%*
%*             Initialization
%*
%************************************************************
\message{jyTeX initialization}

\everyjob{\immediate\write16{--> jyTeX version \fmtversion}%
     \edef\@@jobname{\jobname}%
%     \openin0=\inputpath jysupp
%     \ifeof0
%     \else\closein0
%          \immediate\write16{--> Additional macros loaded from jysupp.tex}%
%          \jyinput jysupp
%     \fi
%     \openin0=\inputpath jylocal
%     \ifeof0
%     \else\closein0
%          \immediate\write16{--> Additional macros loaded from jylocal.tex}%
%          \jyinput jylocal
%     \fi
     \edef\jobname{\@@jobname}%
     \settime
     \openin0=\jobname.lab
     \ifeof0
     \else\closein0
          \immediate\write16{--> Getting labels from file \jobname.lab}%
          \input\jobname.lab
     \fi}

%************** Spacing *************************************

\def\fixedskipslist{%
     \^^\{\topskip}%
     \^^\{\splittopskip}%
     \^^\{\maxdepth}%
     \^^\{\skip\topins}%
     \^^\{\skip\footins}%
     \^^\{\headskip}%
     \^^\{\footskip}}

\def\scalingskipslist{%
     \^^\{\p@renwd}%
     \^^\{\delimitershortfall}%
     \^^\{\nulldelimiterspace}%
     \^^\{\scriptspace}%
     \^^\{\jot}%
     \^^\{\normalbaselineskip}%
     \^^\{\normallineskip}%
     \^^\{\normallineskiplimit}%
     \^^\{\baselineskip}%
     \^^\{\lineskip}%
     \^^\{\lineskiplimit}%
     \^^\{\bigskipamount}%
     \^^\{\medskipamount}%
     \^^\{\smallskipamount}%
     \^^\{\parskip}%
     \^^\{\parindent}%
     \^^\{\abovedisplayskip}%
     \^^\{\belowdisplayskip}%
     \^^\{\abovedisplayshortskip}%
     \^^\{\belowdisplayshortskip}%
     \^^\{\abovechapterskip}%
     \^^\{\belowchapterskip}%
     \^^\{\abovesectionskip}%
     \^^\{\belowsectionskip}%
     \^^\{\abovesubsectionskip}%
     \^^\{\belowsubsectionskip}}

%************** Document layout *****************************

\def\twoupsetup{%                                % setup for twoup style
     \topmargin=.75in
     \leftmargin=.5in
     \vsize=6.9in
     \hsize=4.75in
     \fullhsize=10in
     \let\draft=\relax}

\outputstyle{normal}                             % page style

\def\marginnoteformat{\subscriptsize             % paragraphing of margin notes
     \hsize=1in \baselinestretch=1000 \everypar={}%
     \tolerance=5000 \hbadness=5000 \parskip=0pt \parindent=0pt
     \leftskip=0pt \rightskip=0pt \raggedright}

\head={\ifdraft\normalfonts\it\hfil DRAFT\hfil   % format of headline
     \llap{\number\day\ \monthword\month\ \militarytime}\else\hfil\fi}
\foot={\hfil\normalfonts\numstyle\pagenum\hfil}  % format of footline

\normalbaselineskip=12pt                         % usual \baselineskip
\normallineskip=0pt                              % usual \lineskip
\normallineskiplimit=0pt                         % usual \lineskiplimit
\normalbaselines                                 % set \baselineskip

\topskip=.85\baselineskip \splittopskip=\topskip \headskip=2\baselineskip
\footskip=\headskip

\pagenumstyle{arabic}                            % counter style

\parskip=0pt                                     % no skip between paragraphs
\parindent=20pt                                  % usual \parindent

\baselinestretch=1000                            % set \big-, \med-, \smallskip

%************** Sectioning **********************************

\chapterstyle{left}                              % position of heading
\chapternumstyle{blank}                          % counter style
\def\chapterbreak{\newpage}                      % break before heading
\abovechapterskip=0pt                            % space before heading
\belowchapterskip=1.5\baselineskip               % space after heading
     plus.38\baselineskip minus.38\baselineskip
\def\chapternumformat{\numstyle\chapternum.}     % format of heading counter

\sectionstyle{left}                              % position of heading
\sectionnumstyle{blank}                          % counter style
\def\sectionbreak{\vskip0pt plus4\baselineskip\penalty-100
     \vskip0pt plus-4\baselineskip}              % break before heading
\abovesectionskip=1.5\baselineskip               % space before heading
     plus.38\baselineskip minus.38\baselineskip
\belowsectionskip=\the\baselineskip              % space after heading
     plus.25\baselineskip minus.25\baselineskip
\def\sectionnumformat{%                          % format of heading counter
     \ifblank\chapternumstyle\then\else\numstyle\chapternum.\fi
     \numstyle\sectionnum.}

\subsectionstyle{left}                           % position of heading
\subsectionnumstyle{blank}                       % counter style
\def\subsectionbreak{\vskip0pt plus4\baselineskip\penalty-100
     \vskip0pt plus-4\baselineskip}              % break before heading
\abovesubsectionskip=\the\baselineskip           % space before heading
     plus.25\baselineskip minus.25\baselineskip
\belowsubsectionskip=.75\baselineskip            % space after heading
     plus.19\baselineskip minus.19\baselineskip
\def\subsectionnumformat{%                       % format of heading counter
     \ifblank\chapternumstyle\then\else\numstyle\chapternum.\fi
     \ifblank\sectionnumstyle\then\else\numstyle\sectionnum.\fi
     \numstyle\subsectionnum.}

%************** Footnotes ***********************************

\footnotenumstyle{symbols}                       % counter style
\footnoteskip=0pt                                % jyTeX spacing parameter
\def\footnotenumformat{\numstyle\footnotenum}    % \footnotemark format
\def\footnoteformat{\footnotesize                % paragraphing of text
     \everypar={}\parskip=0pt \parfillskip=0pt plus1fil
     \leftskip=1em \rightskip=0pt
     \spaceskip=0pt \xspaceskip=0pt
     \def\\{\ifhmode\ifnum\lastpenalty=-10000
          \else\hfil\penalty-10000 \fi\fi\ignorespaces}}

%************** Labels **************************************

\def\undefinedlabelformat{$\bullet$}             % mark for undefined label

%************** Equation numbering **************************

\equationnumstyle{arabic}                        % counter style
\subequationnumstyle{blank}                      % counter style
\figurenumstyle{arabic}                          % counter style
\subfigurenumstyle{blank}                        % counter style
\tablenumstyle{arabic}                           % counter style
\subtablenumstyle{blank}                         % counter style

\eqnseriesstyle{alphabetic}                      % sub-counter style for series
\figseriesstyle{alphabetic}                      % sub-counter style for series
\tblseriesstyle{alphabetic}                      % sub-counter style for series

\def\puteqnformat{\hbox{%                        % equation number format
     \ifblank\chapternumstyle\then\else\numstyle\chapternum.\fi
     \ifblank\sectionnumstyle\then\else\numstyle\sectionnum.\fi
     \ifblank\subsectionnumstyle\then\else\numstyle\subsectionnum.\fi
     \numstyle\equationnum
     \numstyle\subequationnum}}
\def\putfigformat{\hbox{%                        % figure number format
     \ifblank\chapternumstyle\then\else\numstyle\chapternum.\fi
     \ifblank\sectionnumstyle\then\else\numstyle\sectionnum.\fi
     \ifblank\subsectionnumstyle\then\else\numstyle\subsectionnum.\fi
     \numstyle\figurenum
     \numstyle\subfigurenum}}
\def\puttblformat{\hbox{%                        % table number format
     \ifblank\chapternumstyle\then\else\numstyle\chapternum.\fi
     \ifblank\sectionnumstyle\then\else\numstyle\sectionnum.\fi
     \ifblank\subsectionnumstyle\then\else\numstyle\subsectionnum.\fi
     \numstyle\tablenum
     \numstyle\subtablenum}}

%************** Reference numbering *************************

\referencestyle{sequential}                      % referencing method
\referencenumstyle{arabic}                       % counter style
\def\putrefformat{\numstyle\referencenum}        % format of reference citation
\def\referencenumformat{\numstyle\referencenum.} % format of number in list
\def\putreferenceformat{%                        % paragraphing of list
     \everypar={\hangindent=1em \hangafter=1 }%
     \def\\{\hfil\break\null\hskip-1em \ignorespaces}%
     \leftskip=\refnumindent\parindent=0pt \interlinepenalty=1000 }

%************** Font initialization *************************

\normalsize

%*****************************************************************************

\def\fmtversion{2.6M (June 1992)}

\catcode`\@=12
% ------------------ End of jytex.tex -----------------

%\input jytex.tex   % available from hep-th
\typesize=10pt \magnification=1200 \baselineskip17truept
%\baselineskip25truept
\footnotenumstyle{arabic} \hsize=6truein\vsize=8.5truein
\input epsf
%\draft
%\leftmargin=1.25in
%\oddleftmargin=.5in
%\evenleftmargin=1.5in
\sectionnumstyle{blank}
\chapternumstyle{blank}
\chapternum=1
\sectionnum=1
\pagenum=0
%\referencestyle{preordered}
% title style follows

\def\begintitle{\pagenumstyle{blank}\parindent=0pt
\begin{narrow}[0.4in]}
\def\endtitle{\end{narrow}\newpage\pagenumstyle{arabic}}

% exercise style follows

\def\beginexercise{\vskip 20truept\parindent=0pt\begin{narrow}[10
truept]}
\def\endexercise{\vskip 10truept\end{narrow}}

% **************    my jyTeX abbreviations   *****************

\def\eql#1{\eqno\eqnlabel{#1}}
\def\ref{\reference}
\def\peq{\puteqn}
\def\pref{\putref}

\def\mgn{\marginnote}
\def\bex{\begin{exercise}}
\def\eex{\end{exercise}}

% *********************** My definitions ************************

\font\open=msbm10 %scaled\magstep1 % For VAX. Borde p195.

 %scaled\magstep1 % For VAX. Borde p195.
%\font\open=msym10 %scaled\magstep1 % For Arbortxt on PC
%\font\opens=msym8 %scaled\magstep1 % For Arbortxt on PC
  % For Arbortxt on PC, and VAX. Borde p199

\def\StretchRtArr#1{{\count255=0\loop\relbar\joinrel\advance\count255 by1
\ifnum\count255<#1\repeat\rightarrow}}
\def\StretchLtArr#1{\,{\leftarrow\!\!\count255=0\loop\relbar
\joinrel\advance\count255 by1\ifnum\count255<#1\repeat}}

\def\StretchLRtArr#1{\,{\leftarrow\!\!\count255=0\loop\relbar\joinrel\advance
\count255 by1\ifnum\count255<#1\repeat\rightarrow\,\,}}

\def\mbox#1{{\leavevmode\hbox{#1}}}

\def\hspace#1{{\phantom{\mbox#1}}}
\def\oZ{\mbox{\open\char90}}

\def\al{\alpha}
\def\bal{{\bmit\alpha}} %in jyTeX
 %in jyTeX
 %in jyTeX
 %in jyTeX
 %in jyTeX
 %in jyTeX
 %in jyTeX
 %in jyTeX
 %in jyTeX
 %in jyTeX
 %in jyTeX
 %in jyTeX
 %in jyTeX
% in jyTeX
% in jyTeX
% in jyTeX
\def\bom{{\bmit\omega}}% in jyTeX
% in jyTeX

\def\ga{\gamma}
\def\de{\delta}
\def\Ga{\Gamma}

\def\la{\lambda}

\def\om{\omega}

\def\Si{\Sigma}

\def\ze{\zeta}

\def\De{\Delta}

\def\det{{\rm det\,}}

\def\sc{{\rm sc }}

\def\zf{$\zeta$--function}
\def\zfs{$\zeta$--functions}

     % Newline

\def\frac#1/#2{\leavevmode\kern.1em
\raise.5ex\hbox{\the\scriptfont0 #1}\kern-.1em/\kern-.15em
\lower.25ex\hbox{\the\scriptfont0 #2}}
\def\sfrac#1/#2{\leavevmode\kern.1em
\raise.5ex\hbox{\the\scriptscriptfont0 #1}\kern-.1em/\kern-.15em
\lower.25ex\hbox{\the\scriptscriptfont0 #2}}

\def\gtorder{\mathrel{\raise.3ex\hbox{$>$}\mkern-14mu
             \lower0.6ex\hbox{$\sim$}}}
\def\ltorder{\mathrel{\raise.3ex\hbox{$<$}\mkern-14mu
             \lower0.6ex\hbox{$\sim$}}}

\def\semidirprod{\rlap{\ss C}\raise1pt\hbox{$\mkern.75mu\times$}}
\def\for{\lower6pt\hbox{$\Big|$}}
\def\fish{\kern-.25em{\phantom{abcde}\over \phantom{abcde}}\kern-.25em}

 %triple
%dot
 %double
%dot
 %double dot
%for small #1

\def\boxit#1{\vbox{\hrule\hbox{\vrule\kern3pt
        \vbox{\kern3pt#1\kern3pt}\kern3pt\vrule}\hrule}}
\def\dalemb#1#2{{\vbox{\hrule height .#2pt
        \hbox{\vrule width.#2pt height#1pt \kern#1pt \vrule
                width.#2pt} \hrule height.#2pt}}}

        %double stroke
\def\frac#1#2{{{#1}\over{#2}}}
 %lower covariant deriv.
 %upper covariant deriv.
 %lower covariant deriv semicolon.
    %lower ordinary  deriv.
    %lower ordinary  deriv comma.

\def\noin{\noindent}

      %Connection
    %Connection'

\def\etc{{\it etc. }}

\def\eg{{\it e.g.}}

 %gives average <#1>
 %gives thermal average <<#1>>
   %gives bracket <#1|#2>
   %gives comma bracket <#1,#2>
 %gives round bracket (#1,#2)
 %gives round bracket (#1,|#2)
 %gives big bracket <#1|#2>
  %gives
%matrix element <#1|#2|#3>

%gives reduced matrix element
%<#1||#2||#3>

\def\3j#1#2#3#4#5#6{\left\lgroup\matrix{#1&#2&#3\cr#4&#5&#6\cr}
\right\rgroup}

\def\m?{\mgn{?}}
% KK's defs

\def\beq{\begin{eqnarray}}
\def\eeq{\end{eqnarray}}

%  *******************  Journal refs **********************

\def\aop#1#2#3{{\it Ann. Phys.} {\bf {#1}} ({#2}) #3}
\def\cjp#1#2#3{{\it Can. J. Phys.} {\bf {#1}} ({#2}) #3}
\def\cmp#1#2#3{{\it Comm. Math. Phys.} {\bf {#1}} ({#2}) #3}

\def\jmp#1#2#3{{\it J. Math. Phys.} {\bf {#1}} ({#2}) #3}
\def\jpa#1#2#3{{\it J. Phys.} {\bf A{#1}} ({#2}) #3}

\def\np#1#2#3{{\it Nucl. Phys.} {\bf B{#1}} ({#2}) #3}

\def\prB#1#2#3{{\it Phys. Rev.} {\bf B{#1}} ({#2}) #3}
\def\prD#1#2#3{{\it Phys. Rev.} {\bf D{#1}} ({#2}) #3}
\def\prl#1#2#3{{\it Phys. Rev. Lett.} {\bf #1} ({#2}) #3}

\def\am#1#2#3{{\it Acta Mathematica} {\bf {#1}} ({#2}) #3}
\def\aim#1#2#3{{\it Adv. in Math.} {\bf {#1}} ({#2}) #3}
\def\ajm#1#2#3{{\it Am. J. Math.} {\bf {#1}} ({#2}) #3}

\def\jpamt#1#2#3{{\it J. Phys.A:Math.Theor.} {\bf{#1}} ({#2}) #3}
\def\jram#1#2#3{{\it J. f. reine u. Angew. Math.} {\bf {#1}} ({#2}) #3}

\def\ma#1#2#3{{\it Math. Ann.} {\bf {#1}} ({#2}) #3}

\def\mz#1#2#3{{\it Math. Zeit.} {\bf {#1}} ({#2}) #3}

\def\plb#1#2#3{{\it Phys. Letts.} {\bf {B#1}} ({#2}) #3}

\def\qjm#1#2#3{{\it Quart. J. Math.} {\bf {#1}} ({#2}) #3}

\def\rmjm#1#2#3{{\it Rocky Mountain J. Math.} {\bf {#1}} ({#2}) #3}

\def\tams#1#2#3{{\it Trans.Am.Math.Soc.} {\bf {#1}} ({#2}) #3}

% *******************   Main text *********************
\begin{title}
\vglue 0.5truein
%\righttext {MUTP/96/23}
%\righttext{hep-th/96}
\vskip15truept
%\leftline{\today}
%\vskip 30truept
\centertext {\Bigfonts \bf Calculation of the } \vskip7truept
\vskip10truept\centertext{\Bigfonts \bf multiplicative anomaly }
 \vskip7truept
\vskip10truept\centertext{\Bigfonts \bf }
 \vskip 20truept
\centertext{J.S.Dowker\footnote{dowkeruk@yahoo.co.uk}} \vskip
7truept \centertext{\it Theory Group,} \centertext{\it School of Physics and Astronomy,}
\centertext{\it The University of Manchester,} \centertext{\it Manchester, England} \vskip
7truept \centertext{}

\vskip 7truept \vskip40truept
\begin{narrow}
The functional determinant multiplicative anomaly, or defect, is more closely investigated
and explicit forms for products of the particular operators, $L-\al_i$,  are produced. I also
present formulae for the defect of products of $L^2-\al_i^2$ in terms of that for just two
factors and discuss the specific cases of the sphere and hemisphere. The difference of
Neumann and Dirchlet quantities on the hemisphere is equal to that for spin--1/2 on the rim.
This is proved generally.
\end{narrow}
\vskip 5truept
%\righttext {August 1996}
\vskip 60truept
%\righttext{Typeset in \jyTeX}
\vfil
\end{title}
\pagenum=0
\newpage

\section{\bf 1. Introduction}

I am interested in the functional determinant of a product of operators. As is well known,
this is, generally, not the product of the determinants, the discrepancy being termed the
`multiplicative anomaly'. An expression was given by Wodzicki in 1987, but the general
phenomenon seems to have been noticed earlier in a physical context, Allen [\pref{Allen2}],
Chodos and Myers, [\pref{Chodos1}]. I do not give later references but just note that
calculations, relevant to some of those here, have been done recently by Cognola, Elizalde
and Zerbini, [\pref{CEZ}] using the Wodzicki approach.

The functional determinant of products of second order elliptic operators (on spheres) has
appeared lately in connection with the AdS/CFT correspondence, \eg\ [\pref{Tseytlin}], but
the possibility of a defect is not considered, perhaps because it could be absorbed in
renormalisation.

Therefore, I thought it might be useful to present some (possibly known) results for defects
but derived in an elementary, spectral way. The operators will be restricted but have
occurred in various contexts. My formulae may not be as general as they could be but they
are, at least, explicit.

\section{\bf 2. The set up and the defects}
I consider a set of operators $D_i$ $i=1,\ldots,k$, and ask for the difference in logdets,
  $$
  \log\det{\textstyle\prod_i}D_i-{\textstyle\sum}_i\log\det D_i\equiv M_k(D_1,\ldots,D_k)
  \eql{ma}
  $$
which I will refer to as a multiplicative anomaly, or, sometimes, defect. All the $D_i$
commute and all the logdets are {\it defined} from the \zf\ of the relevant operator,
assuming this makes sense.

I make a special choice of operator,
  $$
   D_i=L^2-\al_i^2=(L+\al_i)(L-\al_i)\equiv D^{(+)}_i\,D^{(-)}_i
   \eql{dee}
  $$
where the $\al_i$ are constants. $L$ is a pseudo--differential operator, in particular the
square--root of a shifted Laplace--Beltrami operator in $d$--dimensions,
  $$
  L=(-\De_2+c)^{1/2}\,,
  \eql{bee}
  $$
where $c$ is a constant, to be suitably chosen.\footnote{ On the sphere, for a particular
$c$, this operator is strictly Zoll, up to an additional constant. This means the spectrum is
$1,2,3,\ldots$ (with degeneracies).}

The $D_i$ are, therefore, second order. It is a general result that the defect vanishes for
closed  odd--dimensional manifolds.

This choice covers a number of relevant cases. For example, for an $O(n)$ scalar field, the
functional part of the determinant takes the form,
  $$
  \log\det\big[(-\De_2+m_1^2)\ldots(-\De_2+m^2_n)\big]\,,
  $$
and a similar product structure arises in the theory of massless, conformal higher spins on a
sphere, which is a more interesting case and the one I have in mind. In field theory
computations it is commonly assumed that the defect, $M_k$, is zero, even for even $d$.
Although the physical relevance of the defect is doubtful, because of renormalisation, its
computation is still an interesting mathematical question.

To this end I proceed by firstly saying that I do not wish to give any preliminary derivations
beyond those that already exist. This is the reason I have introduced the factorisation of the
second order operator into linear factors since the defects after a complete linear
factorisation have already been evaluated. I therefore define a further anomaly for
complete linear factorisation by,
  $$
  \log\det{\textstyle\prod_i}D_i-{\textstyle\sum}_i\log\det D^{(+)}_i-
  {\textstyle\sum}_i\log\det D^{(-)}_i\equiv L_k(D_1,\ldots,D_k)\,.
  \eql{ma2}
  $$
  In particular, applying this to a single $D_i$ and summing,
  $$
  {\textstyle\sum}_i\log\det D_i-{\textstyle\sum}_i\log\det D^{(+)}_i-
 {\textstyle\sum}_i \log\det D^{(-)}_i= {\textstyle\sum}_i L_1(D_i)
  \eql{ma3}
  $$

Subtracting these   equations, and referring to the definition, (\peq{ma}), I get
  $$
  M_k(D_1,\ldots,D_k)=L_k(D_1,\ldots,D_k)-
 {\textstyle\sum}_{i=1}^k L_1(D_i)
 \eql{em}
  $$
the point now being that the quantities on the right--hand side have been given elsewhere.
A direct spectral means of finding the second term in (\peq{em}) was given in
[\pref{DowGJMS}] and extended to the $k$--fold product  in [\pref{Dowcmp}]. I can,
therefore, just write down the resulting combination in (\peq{em}),
  $$\eqalign{
  M_k(D_1,\ldots,D_k)=&{k-1\over2k}\sum_{r=1}^u\bigg({H(r-1)\,N_{2r}(d)\over r}
  \sum_{j=1}^{k}\al_j^{2r}\bigg)\cr
  &-{1\over2k}\sum_{r=1}^u{1\over r}\sum_{t=1}^{u-r}
  {N_{2r+2t}(d)\over t}\sum_{i<j=1}^{k}\al_i^{2r}\,\al_j^{2t}\,\,.
  }
 \eql{em2}
  $$
The upper limit $u$ equals $d/2$ for even dimensions, and $(d-1)/2$ for odd.\footnote{ If
the manifold is closed only even dimensions are relevant.}

In this formula $H$ is the harmonic series, $H(r)=\sum_{n=1}^r1/n$, $H(0)=0$, and $N$
is the residue at a pole of the \zf\ defined by,
    $$
    Z_d(s)\equiv \sum_m{1\over \la_m^s}\,,
    \eql{zedd}
    $$
where the $\la_m$ are the eigenvalues (with repeats) of the linear operator, $L$,
(\peq{bee}). That is,
    $$
   Z_d(s+r)\rightarrow {N_r(d)\over s}+R_r(d) \quad{\rm
   as}\,\,s\rightarrow 0\,.
  \eql{pole1}
  $$
The $N$s are simply related to the heat--kernel coefficients and are, therefore, locally
computable. $N$ and $R$ will depend on the parameters in the eigenvalues, $\la_m$.

I note the important fact that there are no cubic, or higher powers of $\al^2$ in
(\peq{em2}).

Expression (\peq{em2}) can be implemented easily on a machine, but it does not directly
yield the answer in its simplest factorised form without further manipulation. The
symmetrised sums appearing in (\peq{em2}),
  $$\eqalign{
  \Si_1(k)&=\sum_{j=1}^{k}\al_j^{2r}\cr
  \Si_2(k)&=\sum_{i<j=1}^{k}\al_i^{2r}\,\al_j^{2t}\,,
  }
  $$
 allow some combinatorial rearrangements to be made at an earlier stage.

For short I now write $M_k(D_1,\ldots,D_k)=M(1,\ldots,k)$ and also define $M(i,j)\equiv
M_2(D_i,D_j)$ with $i<j$. The aim is to write everything in terms of  the two operator
anomaly, $M(i,j)$. I spell out the steps.

From (\peq{em2}), defining handy quantities $A$ and $B$,
  $$\eqalign{
  M(i,j)=&{1\over4}\sum_{r=1}^uA(r,d)
  (\al_i^{2r}+\al_j^{2r})
  -{1\over4}\sum_{r=1}^u\sum_{t=1}^{u-r}B(r,t,d)\,
  \al_i^{2r}\,\al_j^{2t}\,.
  }
 \eql{em3}
  $$

Therefore,
  $$\eqalign{
  {2\over k}\sum_{i<j=1}^kM(i,j)=&{1\over2k}\sum_{r=1}^uA(r,d)
  \sum_{i<j=1}^k (\al_i^{2r}+\al_j^{2r})\cr
  &-{1\over2k}\sum_{r=1}^u\sum_{t=1}^{u-r}B(r,t,d)\,
   \sum_{i<j=1}^k\al_i^{2r}\,\al_j^{2t}\,\cr
   &={k-1\over2k}\sum_{r=1}^uA(r,d)
  \sum_{j}^k \al_j^{2r}\cr
  &-{1\over2k}\sum_{r=1}^u\sum_{t=1}^{u-r}B(r,t,d)\,
   \sum_{i<j=1}^k\al_i^{2r}\,\al_j^{2t}\,\cr
   &=M(1,2,\ldots,k)\,,
  }
 \eql{em4}
  $$
which is the desired relation. As a check, since $M(i,j)$ vanishes if $\al^2_i=\al^2_j$ so do
all the anomalies, as required.

If there are repeats amongst the $D_i$, say $D_i$ appears $g_i$ times, then it is easy to
see that the relation is modified to
  $$
   M(1,2,\ldots,k)={1\over g}\sum_{i<j=1}^k (g_i+g_j)\,M(i,j)
   \eql{em5}
  $$
where $g$ is the total degeneracy, $g=\sum_i g_i$. Similar formulae can be found relating
$M_k$ to $M_n$ for any $n$, $1<n<k$.

Equation (\peq{em5}) is a particular example of a general result of Castillo-Garate,
Friedman and M\v{a}ntoiu, [\pref{CFM}], having the same basic definitions but with $D_i$
quite general operators. The proof is an inductive one, and uses general properties of the
Wodzicki expression for the defect.

In the approach of the present paper, the possibility of such a decomposition arises from the
absence of {\it cubic}, and higher, terms in the explicit form of the defect.

\section{\bf3. A special case, the sphere and hemisphere.}

Explicit expressions for the defect can be obtained when the singularity structure of the \zf\
$Z_d$ is known (or equivalently the heat--kernel coefficients). Such is the case for spheres
discussed already in [\pref{Dowcmp,Dowjmp,DowGJMS}]. The motivation for considering
products of operators on the sphere was Branson's expression for the higher derivative
conformal operator (GJMS operator). In that case the parameters, $\al_i$, take a specific
form, but, as we have just shown, the evaluation applies to any distribution.

I find it convenient, and a little more flexible, to consider the sphere as the union of two
hemispheres. To be more precise, the spectrum on the whole sphere is the union of the
Neumann and Dirichlet spectra  on a hemisphere. Then I have access to the hemisphere
values as well.

For this geometry, the eigenvalues of $L$ take the linear form,
$$
  \la_m=a+{\bf m}.\bom
  \eql{lalin}
  $$
where, for the hemisphere, $\bom={\bf 1}_d$.\footnote{ Other choices for $\bom$ can
give other divisions of the sphere. I will not consider these. For integer $\bom$ the
spectrum is strictly Zoll.} Neumann conditions arise when $a=(d-1)/2$, and Dirichlet when
$(d+1)/2$. It is often sufficient to give the expressions for just one of these conditions as the
other is the same up to a sign.

$Z_d$, is then a Barnes \zf\ with explicit pole structure. The residues, $N$ in (\peq{pole1}),
are easily calculated generalised Bernoulli polynomials.

The formula (\peq{em2}) has been evaluated by brute force for various $d$ and $k$. I will
not write out the expressions but rather will appeal to the sum structure (\peq{em4}) which
organises them in a more symmetrical way. Then I need give only $M(i,j)$ for each
dimension. For display purposes, I set $x=\al_i^2,\,y=\al_j^2$ and write $M(x,y)$ for
$M(i,j)$.

Every $M(x,y)$ has the factor $(x-y)^2$ so I put
  $$
  M(x,y)=(x-y)^2\,P(d)
  \eql{defpoly}
  $$
and list $P(d)$, a symmetric polynomial in  $x$ and $ y$. I have put back the dimension.
The Neumann hemisphere values are,
  $$\eqalign{
 & P_N(3)=0,\quad P_N(4)={1\over48},\quad P_N(5)={1\over96},\cr
  &P_N(6)={1\over1920}\left( 2\,(x+y)-5\right),\quad P_N(7)={1\over1920}
  \left( (x+y)-5\right),\cr
  &P_N(8)={1\over1935360}\left( 44\,(x^2+{y}^{2})+56\,{x}\,{y}
  -420\,(x+{y})+777\right),\cr
  &P_N(9)={1\over967680}\left( 11\,(x^2+{y}^{2})+14\,{x}\,{y}
  -168\,(x+{y})+588\right)\,.
  }
  \eql{defect2}
  $$
The Dirichlet values, $P_D(d)$ are the same as these for even dimensions, and opposite in
sign for odd.

For comparison, I also give the corresponding spin--half quantities using the same notation
The non--zero values are,
  $$\eqalign{
 & P_{1/2}(4)={1\over48}\quad P_{1/2}(6)={1\over960}(x+y-5)\cr
  &P_{1/2}(8)={1\over483840}\left( 11(x^2+y^2)+14xy-168(x+y)+588)\right),\cr
  }
  \eql{defect3}
  $$
and I note the holographic--like relation
  $$
  P_N(d)-P_D(d)=P_{1/2}(d-1),\quad\forall\,d\,.
  \eql{hrel}
  $$

I should note here that I am using `mixed' (local) boundary conditions for the spinor field.
These are conformally covariant. Both types (analogous to N and D) give the {\it same}
hemisphere eigenvalue set which is as in (\peq{lalin}) with $a=d/2$. (See the Appendix and
[\pref{Apps}].)

\section{\bf4. Linear factorisation, again}

For the operator (\peq{dee})  the factors occur in `conjugate' pairs, $\pm\al_i$. This means
that all calculations involve $\al_i^2$ from the start, which is somewhat restrictive. In this
section I discuss the slightly more flexible product ($w$ is a constant introduced for
expository convenience),
  $$
  \prod_{i=1}^{2k}(L+w-\al_i)\,,
  \eql{linprod}
  $$
where, to preserve my previous notation, $k$ can be an integer or a half--integer. $L$ can,
in fact, be a fairly general operator, not necessarily as in (\peq{bee}). All I require is that
the corresponding \zf, $Z(s)$,
  $$
  Z(s,w)\equiv\sum_m{1\over(\la_m+w)^s}\,,
  \eql{zed}
  $$
has reasonable properties such as a finite number of single poles (only) and that it is analytic
around $s=0$. In particular I will require the behaviour around the points $s=r$, $r\in
\oZ_+$, which, for a general operator $L$, may, or may not, be poles. I assume therefore
that,
  $$
  Z(s+r,w)\sim {N(r,w)\over s}+R(r,w)\,,\quad s\to 0\,\,\,{\rm and}\,\,r\in\oZ_+\,,
  \eql{zpole}
  $$
where $N(r,w)$ might be zero. The $N$ and $R$ depend on the dimension, $d$, and on the
parameters determining the eigenvalues. I do not show these.

My basic example is the linear function (\peq{lalin}) and $Z$ is then the Barnes function,
already used. See the section 6.

I define the multiplicative anomaly, or rather defect, for the `linear factorisation',
(\peq{linprod}) by (this is just (\peq{ma}) with a different notation),
  $$
  \de(k,w;\bal)\equiv\log\det\prod_{i=1}^{2k}(L+w-\al_i)
  -\sum_{i=1}^{2k}\log\det(L+w-\al_i)\,.
  \eql{defdef}
  $$
 $\bal$ is a $2k$--vector, specifying the operator (\peq{linprod}).

 In the next section I give calculational details of the case of two factors, which is enough to
motivate the general formulae,
$$\eqalign{
  \de(k,w;\bal)=&{2k-1\over2k}\sum_{r=1}^u{H(r-1)\,N(r,w)\over r}
  \bigg(\sum_{j=1}^{2k}\al_j^{r}\bigg)\cr
  &+{1\over2k}\sum_{r=1}^u
  \sum_{t=1}^{u-r}{N(r+t,w)\over r\,t}\,
  \bigg({\sum_{i<j=1}^{2k}\al_i^{r}\,\al_j^{t}}\bigg)
  }
  \eql{del1}
  $$
which correctly vanishes for one factor, $k=1/2$,

The upper limit $u$ equals $[\mu]$ where $\mu$ is the order of the eigenvalue set,
$\la_m$. Most usually, $\mu=d$ for linear operators and $d/2$ for second order ones. The
sums over $r$ and $t$ in (\peq{del1}) restrict to the pole set in $Z(s)$.

Just as in the previous discussion, it is possible to express the general product defect in
terms of the two--factor defect, either in the same manner as before, see (\peq{em4}), or
by rearranging the sums in (\peq{del1}). Then, not surprisingly,
  $$
  \de(k,w;\bal)={1\over k}\sum_{i<j=1}^{2k}\de(1,w;\al_i,\al_j)\,.
  \eql{defdecomp}
  $$

In the next section I expand on the two factor case and apply it to the (hemi)--sphere in the
following section.

\section {\bf 5. Two factors.}
For two factors the \zf\ is,
  $$
\ze(s,w)\equiv\sum_{m}{1\over(  \la_m+w-\al)^s(\la_m+w-\al')^s}\,.
  $$

The method, as used in [\pref{Dowcmp,DowGJMS}], is to binomially expand each bracket
and do the resulting sum over  $m$ using (\peq{zed}). This gives,
  $$
  \ze(s,w)\!\!=\!\sum_{r=0}^\infty\! \sum_{r'=0}^\infty\!\!\al^r\al'^{\,r'}
   {s(s+1)\ldots(s+r-1)s(s-1)\ldots(s+r'-1)\,\over r!\,r'!}Z(2s+r+r'\!\!,w)\,.
  $$
For the determinant of the product (defined by \zfs), I require $\ze'(0,w)$.

To expose the powers of $s$, it is convenient to split off the $r=0$ and $r'=0$ terms.
Written out,
  $$\eqalign{
  \ze(s,w)=&Z(2s,w)+\sum_{r=1}^\infty(\al^r+\al'^{r})
   {s(s+1)\ldots(s+r-1)\over r!}Z(2s+r,w)\cr
   +\sum_{r,r'=1}^\infty& \al^r\al'^{\,r'}
   {s(s+1)\ldots(s+r-1).s(s+1)\ldots(s+r'-1)\,\over r!\,r'!}Z(2s+r+r',w)\cr
   \sim\,\, &Z(2s,w)+\sum_{r=1}^\infty(\al^r+\al'^{r})
   {(s+1)\ldots(s+r-1)\over r!}\bigg({N(r,w)\over2}+sR(r,w)\bigg)+\cr
   &\!\!\!\!\!\!\!\!\!\!\!\sum_{r,r'=1}^\infty\!\!\!\al^r\al'^{\,r'}\!\!
   {(s+1)\ldots(s+r-1).(s+1)\ldots(s+r'-1)\,\over r!\,r'!}\cr
  &\hspace{*********************}\times
  s\bigg(\!\!{N(r\!+\!r',w)\over2}\!+\!sR(r+r',w)\!\!\bigg)\,.
   }
  $$

Therefore
    $$\eqalign{
    \ze'(0,w)=2Z'(0,w)+&\sum_{r=1}^{[\mu]}{\al^r+\al'^{r}\over r}\bigg(
   {1\over2}H(r-1)\,N(r,w)+R(r,w)\bigg)\cr
    &+\sum_{r=[\mu]+1}^\infty{\al^r+\al'^{r}\over r}\,Z(r,w)
     +{1\over2}\!\!\!\sum_{{r,r'=1}\atop r+r'\le[\mu]}^\infty
     {\al^r\al'^{\,r'}\over rr'}\,N(r\!+\!r',w)\,.
   }
   \eql{zd}
   $$
 I do not wish to enlarge on the generality of the operator $L$ and only say that if $\mu$ is
the order of the set of eigenvalues, $\la_m$, $r$ is less than, or equal to, $[\mu]$. For the
eigenvalues (\peq{lalin}), for example, $\mu=d$. $N(r,w)$ is non--zero only if $r\le[\mu]$.

I now convert the infinite series into a finite one in the following, rather roundabout way.
Introduce the `heat--kernel', $K(\tau,w)$, of the operator $L+w$, explicitly
   $$
   K(\tau,w)=\sum_m e^{-(\la_m+w)\,\tau}\,,
   $$
and use the Mellin representation of the \zfs\ to write the infinite sum in (\peq{zd}) as,
using an intermediate, calculational regularisation,
  $$\eqalign{
\sum_{r=d+1}&{\al^r\over r\,\Ga(r)}\int_0^\infty d\tau\tau^{r-1}
K(\tau,w)\cr =&\lim_{s\to0} \int_0^\infty d\tau\,
\left(\exp(\al\tau)-\sum_{r=0}^d{(\al\tau)^{r} \over
r!}\right)\tau^{s-1}K(\tau,w)\cr
=&\lim_{s\to0}{\bigg(\Ga(s)Z(s,w-\al)-\sum_{r=0}^d{\al^r\over r!}\,
\Ga(s+r)Z(s+r,w)\bigg)\,.
}}
  \eql{reg}$$

Since the total quantity in (\peq{reg}) is finite, the individual pole terms that arise in the
$s\to0$ limit must cancel yielding the identity between heat--kernel coefficients,
  $$
Z(0,w-\al)-Z(0,w)=\sum_{r=1}^{[\mu]}{\al^{r}\over r}N(r,w).
\eql{iden1}
  $$

The finite remainder is the required result and equals,
  $$\eqalign{
Z'(0,w-\al)&-Z'(0,w) -\sum_{r=1}^{[\mu]}
{\al^{r}\over r}\,R(r,w)\cr &-\ga\big(Z(0,w-\al)-Z(0,w)\big)
- \sum_{r=1}^{[\mu]} {\al^{r}\over r}\,\psi(r)N(r,w)\,,}
  $$
which, in view of (\peq{iden1}), yields
  $$\eqalign{ & \sum_{r=[\mu]+1}^\infty{\al^r\over
r}\,Z(r,w))\cr &=Z'(0,w-\al)-Z'(0,w))- \sum_{r=1}^{[\mu]} {\al^{r}\over
r}R(r,w)-\sum_{r=1}^{[\mu]} {\al^{r}\over r}
\big(\psi(r)+\ga\big)N(r,w)\cr &= Z'(0,w-\al)-Z'(0,w) -
\sum_{r=1}^{[\mu]} {\al^{r}\over r}\big(R(r,w)+
H(r-1)\,N(r,w)\big)\,,\cr
}
\eql{sum4}
  $$
where $\ga$ is the Euler constant.

Using (\peq{sum4}) in (\peq{zd}) gives my final answer for two factors,,
   $$\eqalign{
    \ze'(0,w)=2Z'(0,w)&+Z'(0,w-\al)-Z'(0,w)+Z'(0,w-\al')-Z'(0,w)
    \cr&+\sum_{r=1}^{[\mu]}{\al^r+\al'^{r}\over r}\bigg(
   {1\over2}H(r-1)\,N(r,w)+R(r,w)\bigg)\cr
    -\sum_{r=1}^{[\mu]}{\al^r+\al'^{r}\over r}&\big(R(r,w)+
H(r-1)\,N(r,w)\big)
     +{1\over2}\!\!\!\sum_{{r,r'=1}\atop r+r'\le[\mu]}^\infty
     {\al^r\al'^{\,r'}\over rr'}\,N(r\!+\!r',w)\cr
     =Z'(0,w-&\al)+Z'(0,w-\al')
    \cr
    -{1\over2}\sum_{r=1}^{[\mu]}{\al^r+\al'^{r}\over r}&\big(
H(r-1)\,N(r,w)\big)
     +{1\over2}\!\!\!\sum_{{r,r'=1}\atop r+r'\le[\mu]}^\infty
     {\al^r\al'^{\,r'}\over rr'}\,N(r\!+\!r',w)\,.
   }
   \eql{zd2}
   $$

 In the special case of $\al'=-\al$
   $$\eqalign{
     \ze'(0,w)=&Z'(0,w-\al)+Z'(0,w+\al)
    \cr
    -\sum_{r=1}^{[\mu]}{(1+(-1)^r)\al^r\over2 r}&
H(r-1)\,N(r,w)
     +{1\over2}\!\!\!\sum_{{r,r'=1}\atop r+r'\le[\mu]}^\infty
     (-1)^{r'}{\al^{r+r'}\over rr'}\,N(r\!+\!r',w)\cr
     &=Z'(0,w-\al)+Z'(0,w+\al)\cr
    -\sum_{\rho=1}^{[\mu/2]}{\al^{2\rho}\over2\rho}
H(2\rho-1)\,&N(2\rho,w)
     +{1\over2}\!\!\!\sum_{{r,r'=1}\atop r+r'\le[\mu]}^\infty{(-1)^r+(-1)^{r'}\over2}
     {\al^{r+r'}\over rr'}\,N(r\!+\!r',w)\,.
   }
   \eql{zd3}
   $$

The last term can be reworked so as to combine with the preceding one.  I give the details
for those who might want to see the nuts and bolts. Setting $r+r'=2\rho$,
  $$\eqalign{
  &{1\over2}\!\!\!\sum_{{r,r'=1}\atop r+r'\le[\mu]}^\infty{(-1)^r+(-1)^{r'}\over2}
      {\al^{r+r'}\over rr'}\,N(r\!+\!r',w)\cr
     &=\sum_{\rho=1}^{[\mu/2]}{\al^{2\rho}\over 2}\,
     N(2\rho,w)\sum_{r'=1}^{2\rho-1}{(-1)^{r'}\over(2\rho-r')\, r'}\cr
     &=\sum_{\rho=1}^{[\mu/2]}{\al^{2\rho}\over 4\rho}\,
     N(2\rho,w)\sum_{r'=1}^{2\rho-1}(-1)^{r'}
     \bigg({1\over r'}+{1\over2\rho-r'}\bigg)\cr
     &=\sum_{\rho=1}^{[\mu/2]}{\al^{2\rho}\over 2\rho}\,
     N(2\rho,w)\sum_{r'=1}^{2\rho-1}(-1)^{r'}
     {1\over r'}\cr
     &={1\over2}\sum_{\rho=1}^{[\mu/2]}{\al^{2\rho}\over 2\rho}\,
     N(2\rho,w)
     \big(H(\rho-1)-2H_O(\rho-1)\big)\,,\cr
     }
  $$
where $H_O(r)\equiv  \sum_{k=0}^r 1/(2k+1)$.

Then, combining this with the penultimate term in (\peq{zd3}), one encounters the
simplification,
  $$\eqalign{
  -H(2\rho-1)+{1\over2}H(\rho-1)-H_O(\rho-1)=-2H_O(\rho-1)
  }
  $$
and the total quantity, from (\peq{zd3}), is, finally,
   $$\eqalign{
      \ze'(0,w)=Z'(0,w-\al)+Z'(0,w+\al)
    -\sum_{\rho=1}^{[\mu]}{\al^{2\rho}\over\rho}H_O(\rho-1)\,N(2\rho,w)\,,
   }
   \eql{zd4}
   $$
where the (spin--zero) logdet defect, $\de_0(d,\al)$, is the (negative of) the final term. This
is the expression given in [\pref{Dowcmp}] derived more simply there by expanding in
$\al^2$ from the start. The equality is gratifying.

The expression (\peq{zd4}) is repeated, and generalised, in [\pref{DoandKi}] where some
history was attempted. A derivation of the basic cancellations, which avoids using the
heat--kernel, as in (\peq{reg}), was  also given.
\section{\bf6. Spherical defects again. Some explicit expressions}
The defect in (\peq{zd4}) is a polynomial in $\al^2$. On the sphere some specific
calculations were performed in [\pref{Dowcmp,Dowjmp}]  and I extend these to the
present situation.

For spherical domains, $\la_m$ is as in (\peq{lalin}) and, as before, $Z(s,w)$ is a Barnes
\zf. The general form of the defect polynomial is easily found since the residues, $N$, of the
Barnes \zf\ are given in terms of generalised Bernoulli polynomials, which are readily found.

Without loss of generality, I can now set $w=0$ since it only adds to the constant $a$.

Machine evaluation of the last term in (\peq{zd4}) yields for the hemisphere spin--zero
$N$--defect, $\de_0(d,\al)\big|_N$,
  $$\eqalign{
  &d=2,\quad {\alpha}^{2}        \cr
  &d=3,\quad  \frac{{\alpha}^{2}}{2}     \cr
  &d=4,\quad    \frac{{\alpha}^{4}}{9}-  \frac{{\alpha}^{2}}{24}       \cr
  &d=5,\quad   \frac{{\alpha}^{4}}{18}  -\frac{{\alpha}^{2}}{12}
      \cr
  &d=6,\quad  \frac{23\,{\alpha}^{6}}{5400}
  -\frac{{\alpha}^{4}}{72}
  +\frac{3\,{\alpha}^{2}}{640}\cr
  &d=7,\quad  \frac{23\,{\alpha}^{6}}{10800}
  =\frac{{\alpha}^{4}}{72}+\frac{{\alpha}^{2}}{60}    \cr
  &d=8,\quad  \frac{11\,{\alpha}^{8}}{132300}
  -\frac{23\,{\alpha}^{6}}{25920}
  +\frac{37\,{\alpha}^{4}}{17280}-\frac{5\,{\alpha}^{2}}{7168}       \cr
  &d=9,\quad  \frac{11\,{\alpha}^{8}}{264600}-\frac{23\,{\alpha}^{6}}{32400}
  +\frac{7\,{\alpha}^{4}}{2160}-\frac{{\alpha}^{2}}{280}  \cr
  }
  \eql{def}
  $$

The $D$--hemisphere values, $\de_0(d,\al)\big|_D$, are the same as the $N$ ones in even
dimensions and opposite in sign for odd. Hence, by addition, the defect is zero on odd--
dimensional spheres and twice the values in (\peq{def}) for even. The even values (on the
sphere) have been calculated recently by Cognola, Elizalde and Zerbini, [\pref{CEZ}], using
a different method involving the Wodzicki residue. Our overlapping results agree.

Figs. 1 and 2 plot the defect on some odd and even hemispheres. I note that, in each case, the
roots approximately coincide, more exactly with increasing dimension when the periodicity tends to unity.

\epsfxsize=5truein \epsfbox{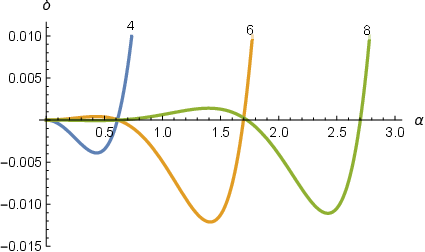}

Fig.1. {\smallfonts Scalar Neumann defect, $\de$, on even hemispheres, dimensions 4,6,8, \break
\hspace{************} plotted against the spectral parameter, $\al$.}
 \vglue 20truept
 \vskip  5truept

\epsfxsize=5truein \epsfbox{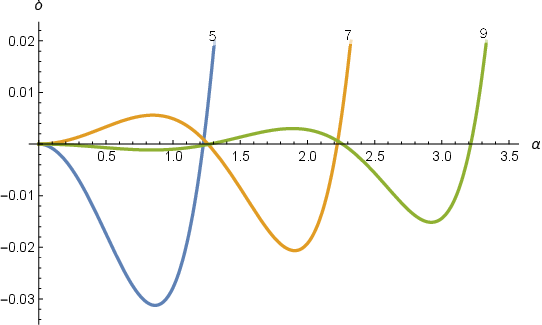}

Fig.2. {\smallfonts Same as Fig.1 but for odd hemispheres, dimensions 5,7,9. }
 \vglue 20truept
 \vskip  5truept

For variety I also give the spin--half hemisphere polynomial defects, $\de_{1/2}(d,\al)$.
The two boundary conditions now give the same values. This means that, since the defect is
zero on odd--dimensional closed spaces, the {\it hemisphere} values are also zero for odd
dimensions, as calculation confirms.\footnote{ This can be inderstood from the fact that the local spinor boundary conditions are a combination of N and D.} The even values for $\de_{1/2}(d,\al)$ are, up to spin
degeneracy,

 $$\eqalign{
  &d=2, \quad    \al^2   \cr
  &d=4,\quad  \frac{{\alpha}^{4}}{9}-\frac{{\alpha}^{2}}{6}    \cr
  &d=6,\quad  \frac{23\,{\alpha}^{6}}{5400}-
  \frac{{\alpha}^{4}}{36}+\frac{{\alpha}^{2}}{30}\cr
  &d=8,\quad  \frac{11\,{\alpha}^{8}}{132300}
  -\frac{23\,{\alpha}^{6}}{16200}+\frac{7\,{\alpha}^{4}}{1080}
  -\frac{{\alpha}^{2}}{140}    \cr
  &d=10,\quad   \frac{563\,{\alpha}^{10}}{571536000}
  -\frac{11\,{\alpha}^{8}}{317520}+\frac{299\,{\alpha}^{6}}{777600}
  -\frac{41\,{\alpha}^{4}}{27216}+\frac{{\alpha}^{2}}{630}   \,.               \cr
  }
  \eql{def2}
  $$

%A graph follows,

%\epsfxsize=5truein \epsfbox{multanom3.eps}

I note the linear counterpart of the relation, (\peq{hrel}),
  $$
  \de_0(d,\al)\big|_N-\de_0(d,\al)\big|_D=\de_{1/2}(d-1,\al)\,,\quad\forall\, d\,,
  \eql{hrel2}
  $$
which connects bulk and boundary quantities. This relation complements one in
[\pref{Dowbfe}] which identifies the difference in the $N$-- and $D$--hemisphere scalar
logdets, for even $d$, with the Dirac (squared) logdet on the odd boundary.

It is possible to give a simple spectral argument that encompasses all these relations. I
outline it in an appendix.

\section{\bf 7. The Wodzicki approach}

Wodzicki has computed the defect for a \zf\ of the form
   $$
   \sum_{n=1}^\infty{P(n)\over \prod_i(n-a_i)^s}
   $$
where $P(n)$ is a polynomial. (See Kassel, [\pref{Kassel}], \S6.6.) This is equivalent to the
hemisphere case if $P(n)$ is the relevant, Barnes degeneracy. The usual way of dealing with
this form is to write the single factor \zf\ (which is a Barnes one) as a sum of Hurwitz
functions. The same technique holds for a general $P(n)$ and the defect can be evaluated by
the present method. The details might be presented at a later time.

\section{\bf8. Conclusion}

The defect for a product of second order scalar operators compared to a sum of the
operators is given in terms of the defect for just two factors. Some examples are given. The
defect for a product of `linear' factors is also given which reduces to a known polynomial for
two conjugate factors. Applied to hemispheres this yields explicit polynomials whose plots
reveal the unexpected feature of (approximately) coincident roots which, at the moment, I cannot explain,
except that it involves the infinite dimensional limit. I also briefly give the corresponding
spin--half defects which turn out to be related to the scalar defects in one dimension higher.

\section{\bf \it Added note Sept.2023} 

I thank Danilo Diaz for communications and pointing out a sign error which has now been corrected. An alternative Shintani-type approach with a relation to a Casimir energy has recently appeared, [\pref{ABDZ}], agreeing with the updated results here.

\section{\bf Appendix A bulk--boundary relation}
I now show that the relations (\peq{hrel}) and (\peq{hrel2}) follow from a very basic
property of Barnes \zfs.

All I need are the eigenvalues on the hemisphere. These are, in $d$ dimensions,
  $$
  \la^N_m(d)={d-1\over2}+{\bf m.1}_d\,,\quad \la^D_m(d)=\la^N_m(d)+1\,,
  \eql{eigs}
  $$
for Neumann and Dirichlet conditions respectively. ${\bf m}$ ranges over all the
non--negative integers. The spin--half eigenvalues are,
  $$
  \la^{1/2}_m(d)=\la^N_m(d)+{1\over2}\,,
  $$
up to spin degeneracy, which I ignore. It is easy to  see that,
  $$
  \{\la^N_m(d)\}-\{\la^D_m(d)\}\equiv\{\la^{1/2}_m(d-1)\}\,,
  $$
as eigenvalue sets.

Assembling the eigenvalues into Barnes \zfs, $\ze_d(s,a\mid\bom)$, this statement leads to
an example of the general recursion relation, due to Barnes,
  $$
  \ze_d(s,a+\om_1\mid \bom)-\ze_d(s,a\mid \bom)=-\ze_{d-1}(s,a\mid\om_2,
  \ldots,\om_d)\,,
  \eql{brc}
  $$
 which specialises here to,
   $$
   \ze^N(s,d)-\ze^D(s,d)=\ze^{1/2}(s,d-1)\,.
   \eql{bbrel}
   $$
Thus all quantities derivable from the \zf\ such as the conformal and multiplicative
anomalies, logdets, free energies, vacuum energies \etc\ , will obey this bulk--boundary
relation in which the bulk quantity is the difference of two boundary conditions. Examples
have appeared in the main body of this paper and also in [\pref{DowGJMS,Dowbfe}].

This result carries through to product operators. It also extends to other values of the
parameters, $\bom$, for example to $\bom=(q,{\bf 1}_{d-1})$ which gives a lune of angle
$\pi/q$ instead of the hemisphere. The right--hand side of (\peq{brc}) is then independent
of $q$, which corresponds to the fact that the boundary of any lune is a full sphere,
metrically. This geometry is useful in discussions of R\'enyi and entanglement entropies.

 \vglue 20truept

 \noin{\bf References.} \vskip5truept
\begin{putreferences}
  \ref{ABDZ}{Aros,R, Bugini,F., Diaz, D.E. and Zuniga ,B., {\it Multiplicative Anomaly matches Casimir  Energy for GJMS Operators on Spheres}, arXiv:2309.04471.}
  \ref{Apps}{Apps,J.S. Thesis (University of Manchester, 1995).}
\ref{Dowbfe}{Dowker,J.S. {\it The boundary F-theorem for free fields}, ArXiv:1407.5909.}
    \ref{DandKi}{Dowker,J.S. and Kirsten, K. {\it Comm. in Anal. and Geom.}
    {\bf7} (1999) 641.}
    \ref{Kassel}{Kassel,C. Seminaire Bourbaki  n. 708 (1988-89) 199.}
 \ref{DoandKi} {Dowker.J.S. and Kirsten, K. {\it Analysis and Appl.}
   {\bf 3} (2005) 45.}
   \ref{CEZ}{Cognola,G.,Elizalde,E. and Zerbini,S. {\it Functional Determinant of the Massive
   Laplace operator and the Multiplicative Anomaly},ArXiv:1408.1766.}
 \ref{CFM}{Castillo-Garate,V.,Friedman,E. and M\v{a}ntoiu,M. {\it The multiplicative
 anomaly of three or more commuting elliptic operstors}, ArXiv:1211.4117.}
 \ref{Allen2}{Allen,B. PhD Thesis, University of Cambridge, 1984.}
 \ref{Chodos1}{Chodos,A. and Myers,E. \aop{156}{1984}{412}.}
     \ref{Guillarmou}{Guillarmou,C. \ajm{131}{2009}{1359}.}
     \ref{Tseytlin}{Tseytlin,A.A. \np{877}{2013}{598}.}
     \ref{BaandDu}{Basar,G and Dunne,G.V. \jpa {43}{2010}{072002}.}
     \ref{AaandD}{Aros,R. and Diaz,D.E. {\it Determinant and Weyl anomaly of
     Dirac operator: a holographic derivation}, ArXiv:1111.1463.}
     \ref{BaandS}{B\"ar,C. and Schopka,S. The Dirac determinant of spherical
     space forms,\break {\it Geom.Anal. and Nonlinear PDEs} (Springer, Berlin, 2003).}
     \ref{QandC}{J.R.Quine and J.Choi, \rmjm {26}{1996}{719-729}.}
  \ref{QandC2}{J.R.Quine and J.Choi, Zeta regularized products and functional
  determinants on spheres \rmjm {26}{1996}{719-729}.}
  \ref{moller}{M{\o}ller,N.M. \ma {343}{2009}{35}.}
     \ref{KKY}{Kanemitsu,S., Kumagai,H. and Yoshimoto,M. {\it The Ramanujan
    J.} {\bf 5}(2001)5.}
     \ref{KandB}{Kamela,M. and Burgess,C.P. \cjp{77}{1999}{85}.}
     \ref{Norlund}{N\"orlund,N.E. \am{43}{1922}{121}.}
     \ref{Dow20}{Dowker,J.S. \jmp{35}{1994}{6076}.}
     \ref{DowGJMSO}{Dowker,J.S. {\it Numerical evaluation of spherical GJMS operators}
     ArXiv: \break 1309.2873.}
     \ref{Adamchik1}{V.S.Adamchik, Polygamma functions of negative order
   {\it J. Comp. Appl. Math.} {\bf 100} (1998) 191-199.}
   \ref{Adamchik}{V.S.Adamchik, {\it Contributions to the theory of the
   Barnes function}, ArXiv: math. CA/0308086.}
     \ref{Vardi}{Vardi,I. {\it SIAM J.Math.Anal.} {\bf 19} (1988) 493.}
    \ref{Onodera}{Onodera,K. \aim{224}{2010}{895}.}
    \ref{KKY}{Kanemitsu,S., Kumagai,H. and Yoshimoto,M. {\it The Ramanujan
    J.} {\bf 5}(2001)5.}
     \ref{Juhl}{Juhl,A. {\it On conformally covariant powers of the Laplacian}
     ArXiv: 0905.3992}
     \ref{Elphinstone}{Elphinstone,H.W. \qjm {2}{1858}{252}.}
       \ref{Branson}{Branson,T.P. \tams{347} {1995}{3671}.}
      \ref{Bromwich}{Bromwich, T.J.I'A. {\it Infinite Series},
  (Macmillan, London, 1926).}
   \ref{Loney}{Loney, S.L. {\it Plane Trigonometry} (CUP, Cambridge, 1893).}
    \ref{Hertzberg}{Hertzberg,M.P. \jpa{46}{2013}{015402}.}
     \ref{CaandW}{Callan,C.G. and Wilczek,F. \plb{333}{1994}{55}.}
    \ref{CaandH}{Casini,H. and Huerta,M. \plb{694}{2010}{167}.}
    \ref{Lindelof}{Lindel\"of,E. {\it Le Calcul des Residues} (Gauthier--Villars, Paris,1904).}
    \ref{CaandC}{Calabrese,P. and Cardy,J. {\it J.Stat.Phys.} {\bf 0406} (2004) 002.}
    \ref{MFS}{Metlitski,M.A., Fuertes,C.A. and Sachdev,S. \prB{80}{2009}{115122}.}
    \ref{Gromes}{Gromes, D. \mz{94}{1966}{110}.}
    \ref{Pockels}{Pockels, F. {\it \"Uber die Differentialgleichung $\De
  u+k^2u=0$} (Teubner, Leipzig. 1891).}
   \ref{Diaz}{Diaz,D.E. JHEP {\bf 0807} (2008) 103.}
    \ref{DandD}{Diaz,D.E. and Dorn,H. JHEP {\bf 0705} (2007) 46.}
  \ref{Minak}{Minakshisundaram,S. {\it J. Ind. Math. Soc.} {\bf 13} (1949) 41.}
    \ref{CaandWe}{Candelas,P. and Weinberg,S. \np{237}{1984}{397}.}
     \ref{Chodos1}{Chodos,A. and Myers,E. \aop{156}{1984}{412}.}
     \ref{ChandD}{Chang,P. and Dowker,J.S. \np{395}{1993}{407}.}
    \ref{LMS}{Lewkowycz,A., Myers,R.C. and Smolkin,M. {\it Observations on
    entanglement entropy in massive QFTs.} ArXiv:1210.6858.}
    \ref{Bierens}{Bierens de Haan,D. {\it Nouvelles tables d'int\'egrales
  d\'efinies}, (P.Engels, Leiden, 1867).}
    \ref{DowGJMS}{Dowker,J.S.  \jpa{44}{2011}{115402}.}
    \ref{Dowren}{Dowker,J.S. \jpamt {46}{2013}{2254}.}
    \ref{Doweven}{Dowker,J.S. {\it Entanglement entropy on even spheres.}
    ArXiv:1009.3854.}
     \ref{Dowodd}{Dowker,J.S. {\it Entanglement entropy on odd spheres.}
     ArXiv:1012.1548.}
    \ref{DeWitt}{DeWitt,B.S. {\it Quantum gravity: the new synthesis} in
    {\it General Relativity} edited by S.W.Hawking and W.Israel (CUP,Cambridge,1979).}
    \ref{Nielsen}{Nielsen,N. {\it Handbuch der Theorie von Gammafunktion}
    (Teubner,Leipzig,1906).}
    \ref{KPSS}{Klebanov,I.R., Pufu,S.S., Sachdev,S. and Saddi,B.R.
    {\it JHEP} 1204 (2012) 074.}
    \ref{KPS2}{Klebanov,I.R., Pufu,S.S. and Safdi,B.R. {\it F-Theorem without
    Supersymmetry} 1105.4598.}
    \ref{KNPS}{Klebanov,I.R., Nishioka,T, Pufu,S.S. and Safdi,B.R. {\it Is Renormalized
     Entanglement Entropy Stationary at RG Fixed Points?} 1207.3360.}
    \ref{Stern}{Stern,W. \jram {79}{1875}{67}.}
    \ref{Gregory}{Gregory, D.F. {\it Examples of the processes of the Differential
    and Integral Calculus} 2nd. Edn (Deighton,Cambridge,1847).}
    \ref{MyandS}{Myers,R.C. and Sinha, A. \prD{82}{2010}{046006}.}
   \ref{RyandT}{Ryu,S. and Takayanagi,T. JHEP {\bf 0608}(2006)045.}
    \ref{Dowcmp}{Dowker,J.S. \cmp{162}{1994}{633}.}
     \ref{Dowjmp}{Dowker,J.S. \jmp{35}{1994}{4989}.}
      \ref{Dowhyp}{Dowker,J.S. \jpa{43}{2010}{445402}.}
       \ref{HandW}{Hertzberg,M.P. and Wilczek,F. \prl{106}{2011}{050404}.}
      \ref{dowkerfp}{Dowker,J.S.\prD{50}{1994}{6369}.}
       \ref{Fursaev}{Fursaev,D.V. \plb{334}{1994}{53}.}
       \ref{Barnesa}{Barnes,E.W. {\it Trans. Camb. Phil. Soc.} {\bf 19} (1903) 374.}
  \ref{Barnesb}{Barnes,E.W. {\it Trans. Camb. Phil. Soc.}{\bf 19} (1903) 426.}
\end{putreferences}

\bye